\documentclass[bibyear]{aa}
\usepackage[varg]{txfonts}
\usepackage{graphicx}
\usepackage{hyperref}
\makeatletter
 \renewcommand\@biblabel[1]{}
\makeatother

\begin{document}

   \title{Infrequent visitors of the Kozai kind: the dynamical lives of 
          2012~FC$_{71}$, 2014~EK$_{24}$, 2014~QD$_{364}$, and 2014~UR
          \thanks{Figures \ref{controlfc71}, \ref{fEWfc71T}, \ref{fEWek24T}, 
                  \ref{fEWqd364T}, \ref{fEWurT}, \ref{controlxx39}, Table 
                  \ref{XX39}, and Appendix A are available in electronic form 
                  at http://www.aanda.org}
         }
   \author{C. de la Fuente Marcos
           \and
           R. de la Fuente Marcos}
   \authorrunning{C. de la Fuente Marcos \and R. de la Fuente Marcos}
   \titlerunning{Infrequent visitors: 2012~FC$_{71}$, 2014~EK$_{24}$, 
                                      2014~QD$_{364}$, and 2014~UR
                 }
   \offprints{C. de la Fuente Marcos, \email{carlosdlfmarcos@gmail.com}
              }
   \institute{Apartado de Correos 3413, E-28080 Madrid, Spain}
   \date{Received 16 March 2015 / Accepted 2 June 2015}

   \abstract
      {Asteroids with semi-major axes very close to that of a host planet can avoid node 
       crossings when their nodal points are at perihelion and at aphelion. This layout 
       protects the asteroids from close encounters, and eventual collisions, with the 
       host planet. 
       }
      {Here, we study the short-term dynamical evolution of four recently discovered 
       near-Earth asteroids (NEAs) ---2012~FC$_{71}$, 2014~EK$_{24}$, 2014~QD$_{364}$, 
       and 2014~UR--- that follow very Earth-like orbits.
       }
      {Our analysis is based on results of direct $N$-body calculations that use the 
       most updated ephemerides and include perturbations from the eight major planets, 
       the Moon, the barycentre of the Pluto-Charon system, and the three largest 
       asteroids.
       }
      {These four NEAs exhibit an orbital evolution unlike any other known near-Earth 
       object (NEO). Beyond horseshoe, tadpole, or quasi-satellite trajectories, they 
       follow co-orbital passing orbits relative to the Earth within the Kozai domain. 
       Our calculations show that secular interactions induce librations of their 
       relative argument of perihelion with respect to our planet but also to Venus, 
       Mars, and Jupiter. Secular chaos is also present. The size of this transient 
       population is probably large.
       }
      {Although some of these NEAs can remain orbitally stable for many thousands of 
       years, their secular dynamics are substantially more complicated than commonly 
       thought and cannot be properly described within the framework of the three-body 
       problem alone owing to the overlapping of multiple secular resonances. Objects 
       in this group are amongst the most atypical NEOs regarding favourable 
       visibility windows because these are separated in time by many decades or even 
       several centuries.
       }


   \maketitle

   \section{Introduction}
      In a seminal paper, Milani et al. (1989) pointed out that for asteroids with semi-major axes very close to that of a host 
      planet, node crossings can be avoided when the nodal points are at perihelion and at aphelion, i.e. at $\omega\simeq0$\degr 
      and $\omega\simeq180$\degr. This arrangement protects the asteroids from close encounters and possible collisions with the 
      host planet, and corresponds to one of the variants of a more general mechanism known as the Kozai resonance (Kozai 1962). 
      This form of the Kozai resonance is particularly important for objects moving in low-eccentricity, low-inclination orbits. 

      This special, dynamically cold case was explored by Michel \& Thomas (1996) who confirmed that, at low inclinations, 
      the argument of perihelion of near-Earth objects (NEOs) can librate around either 0\degr or 180\degr. These authors found 
      that, in the future, 4660 Nereus (1982 DB) will stay in the Kozai resonance for almost 2$\times$10$^{5}$ yr with $\omega$ 
      librating around 180\degr and its inclination oscillating between 5\degr and 11\degr. A similar behaviour was found for the 
      future dynamical evolution of 4034 Vishnu (1986 PA). Such orbital architecture indeed provides a protection mechanism 
      against close encounters with our planet. 

      This interesting topic received additional attention from Namouni (1999) who predicted the existence of minor bodies 
      following passing orbits with small Jacobi constants but still moving in unison with their host planets. Namouni labelled 
      this orbital regime {\it \emph{``the Kozai domain''}} as it is characterised by a libration about 0\degr of the variation of 
      the relative argument of perihelion ($\dot{\omega}_{\rm r}$) with the variation of the relative longitude of the ascending 
      node being negative ($\dot{\Omega}_{\rm r} < 0$). No such objects were found until de la Fuente Marcos \& de la Fuente 
      Marcos (2013) identified 2012~FC$_{71}$ as a Kozai librator that is currently following an Earth-like passing orbit with a 
      small Jacobi constant. This object is locked in a Kozai resonance with $\omega$ librating around 0\degr. In sharp contrast 
      with horseshoe librators, NEOs trapped in the Kozai regime exhibit a very slow orbital evolution and may remain relatively 
      unperturbed for hundreds of thousands of years (Michel \& Thomas 1996; Gronchi \& Milani 1999; de la Fuente Marcos \& de la 
      Fuente Marcos 2013).

      Here we present three recently discovered NEOs ---2014~EK$_{24}$, 2014~QD$_{364}$, and 2014~UR--- that are following 
      dynamically cold, passing orbits with small Jacobi constants with respect to the Earth; i.e. they are co-orbital to our 
      planet. Like in the case of 2012~FC$_{71}$, the argument of perihelion of 2014~EK$_{24}$ and 2014~QD$_{364}$ currently 
      librates around 0\degr. On the other hand, a libration of $\omega$ about 180\degr is observed for 2014~UR. This paper is 
      organised as follows. In Section 2, we briefly outline our numerical model. Section 3 reviews the case of 2012~FC$_{71}$. 
      The dynamics of 2014~EK$_{24}$ is studied in Section 4. Sections 5 and 6 focus on 2014~QD$_{364}$ and 2014~UR, respectively. 
      In Section 7, we show the multi-planet $e_{\rm r} \omega_{\rm r}$-portrait for these objects. Our results are discussed in 
      Section 8. Artificial interlopers are considered in Section 9. Section 10 summarises our conclusions.

   \section{Numerical model}
      Our physical model takes into account the perturbations by eight major planets and treats the Earth--Moon system as two 
      separate objects; it also includes the barycentre of the dwarf planet Pluto--Charon system and the three largest asteroids 
      (for further details, see de la Fuente Marcos \& de la Fuente Marcos 2012). Initial conditions (positions and velocities in 
      the barycentre of the solar system) have been obtained from the Jet Propulsion Laboratory (JPL) HORIZONS system (Giorgini et 
      al. 1996; Standish 1998) and they are referred to as the JD 2457000.5 epoch (2014-December-9.0), which is the $t$ = 0 
      instant in our figures. The numerical integrations discussed in this paper have been performed using the Hermite scheme 
      described by Makino (1991) and implemented by Aarseth (2003). The standard version of this direct $N$-body code is publicly 
      available from the IoA web site.\footnote{\url{http://www.ast.cam.ac.uk/~sverre/web/pages/nbody.htm}} 

      In our calculations, relativistic and oblateness terms and the role of the Yarkovsky and 
      Yarkovsky--O'Keefe--Radzievskii--Paddack (YORP) effects (see e.g. Bottke et al. 2006) have been ignored. Neglecting these 
      effects has no impact on the evaluation of the present dynamical status of the objects discussed here, but may affect both
      the reconstruction of their dynamical past and any predictions made regarding their future evolution. Accurate modelling of 
      the Yarkovsky force requires relatively precise knowledge of the physical properties of the objects affected, which is not 
      the case here. In addition to -- and because of -- the co-orbital status of these bodies, the orbital drifts induced by 
      non-gravitational forces may be smaller than those observed in standard NEOs. In general, oblateness terms induce 
      circulation of the relative argument of perihelion, which can dominate over the objects interaction and terminate the 
      librations of the relative argument of perihelion (Namouni 1999). However, for the objects studied here this contribution 
      could be negligible owing to the low orbital eccentricity and the absence of very close encounters with our planet. 

      In addition to the integrations performed making use of the nominal orbital parameters in Table \ref{elements}, we have 
      computed 50 control simulations for each object with sets of orbital elements obtained from the nominal ones within the 
      quoted uncertainties and assuming Gaussian distributions for them (up to 9$\sigma$ in the cases of 2012~FC$_{71}$ and 
      2014~QD$_{364}$, and 6$\sigma$ for 2014~EK$_{24}$ and 2014~UR). For the sake of clarity, only a few characteristic orbits 
      are shown in the figures. Additional sets of 100 shorter control simulations are discussed in Appendix A (see Figs. 
      \ref{STEfc71} to \ref{STEur}). Relative errors in the total energy at the end of the calculations are $< 1 \times 10^{-15}$. 
      The corresponding error in the total angular momentum is several orders of magnitude smaller. 
%
%------------------------------------------------------------------------------------------------------------------------- TABLE I
%----------------------------------------------------------- Orbital elements asteroids 2012 FC71, 2014 EK24, 2014 QD364 & 2014 UR
%
      \begin{table*}
         \centering
         \fontsize{8}{11pt}\selectfont
         \tabcolsep 0.15truecm
         \caption{\label{elements}Heliocentric Keplerian orbital elements of asteroids 2012~FC$_{71}$, 2014~EK$_{24}$, 
                  2014~QD$_{364}$, and 2014~UR.  
                 }
         \begin{tabular}{cccccc}
            \hline\hline
             Parameter                                             &   &   2012~FC$_{71}$        &   2014~EK$_{24}$               
                                                                   &   2014~QD$_{364}$       &   2014~UR                   \\ 
            \hline
             Semi-major axis, $a$ (AU)                         & = &   0.988482$\pm$0.000009 &   1.00432198$\pm$0.00000002   
                                                                   &   0.989075$\pm$0.000003 &   0.99905206$\pm$0.00000013 \\
             Eccentricity, $e$                                 & = &   0.0880$\pm$0.0002     &   0.0723275$\pm$0.0000002    
                                                                   &   0.04122$\pm$0.00002   &   0.01324277$\pm$0.00000010 \\
             Inclination, $i$ (\degr)                          & = &   4.943$\pm$0.010       &   4.72207$\pm$0.00002       
                                                                   &   3.971$\pm$0.002       &   8.226747$\pm$0.000005     \\
             Longitude of the ascending node, $\Omega$ (\degr) & = &  38.1843$\pm$0.0013     & 341.915256$\pm$0.000008      
                                                                   & 158.2446$\pm$0.0007     &  25.3316268$\pm$0.0000014   \\
             Argument of perihelion, $\omega$ (\degr)          & = & 348.04$\pm$0.02         &  62.44932$\pm$0.00005       
                                                                   &  28.359$\pm$0.014       & 247.5057$\pm$0.0005         \\
             Mean anomaly, $M$ (\degr)                         & = &  67.56$\pm$0.04         &  28.13492$\pm$0.00003        
                                                                   & 248.999$\pm$0.014       & 161.3918$\pm$0.0005         \\
             Perihelion, $q$ (AU)                              & = &   0.9015$\pm$0.0002     &   0.9316819$\pm$0.0000002   
                                                                   &   0.94830$\pm$0.00002   &   0.9858218$\pm$0.0000002   \\
             Aphelion, $Q$ (AU)                                & = &   1.075474$\pm$0.000010 &   1.07696204$\pm$0.00000002  
                                                                   &   1.029848$\pm$0.000003 &   1.01228228$\pm$0.00000013 \\
             Absolute magnitude, $H$ (mag)                     & = &  25.2                   &  23.2                          
                                                                   &  27.2                   &  26.6                       \\
            \hline
         \end{tabular}
         \tablefoot{Values include the 1$\sigma$ uncertainty. The orbits are computed at Epoch JD 2457000.5, which corresponds to 
                    0:00 UT on 2014 December 9 (J2000.0 ecliptic and equinox). Source: JPL Small-Body Database.
                   }
      \end{table*}
%
%---------------------------------------------------------------------------------------------------------------------------------
%

      It may be argued that presenting these objects here could be premature because some of their orbits are still not well known 
      (see Table \ref{elements} and Sections 3--6), in particular those of 2012~FC$_{71}$ and 2014~QD$_{364}$. However,  not all 
      orbital solutions with relatively short data-arc spans are ``born equal''. It is not the same short data-arc for an object 
      that moves in a low-inclination orbit that crosses the paths of all the planets from Jupiter to Mercury as another similarly 
      short data-arc for an object that is only directly perturbed by the Earth--Moon system. In general, these are peculiar, very 
      stable minor bodies with very long synodic periods. The synodic period of an object relative to the Earth is the time 
      interval for the object to return to the same position as seen from our planet (e.g. Green 1985). The synodic period, $S$, 
      is given by $S^{-1} = |T^{-1} - T_{\rm E}^{-1}|$, where $T$ and $T_{\rm E}$ are the orbital periods of the object and the 
      Earth, respectively. For a given object (natural or artificial), it is the characteristic timescale between favourable 
      visibility windows. The relative mean longitude of these objects does not librate as it does in the case of quasi-satellite, 
      tadpole, or horseshoe orbits because $\dot{\omega}_{\rm r}$ librates about 0\degr. Therefore, if an object is only observed 
      a few times after discovery, we have to wait an entire synodic period (not just a fraction of it) to be able to observe it 
      again. In the case of 2012~FC$_{71}$, the waiting time is nearly 56 yr and for 2014~EK$_{24}$ could be as high as 166 yr. 
      For objects in this group, it may take many decades or even several centuries to recover them. No other group of NEOs, 
      except for horseshoe librators, is so dramatically restricted regarding favourable visibility windows.   

   \section{Asteroid 2012~FC$_{71}$, an Aten Kozai librator}
      Asteroid 2012~FC$_{71}$ was discovered on 2012 March 31 by A.~Boattini observing with the Steward Observatory 0.9 m 
      Spacewatch telescope at Kitt Peak (Scotti et al. 2012). It is a small object with $H$ = 25.2, which translates into a 
      diameter in the range 20--60 m for an assumed albedo of 0.20--0.04. The orbital elements of 2012~FC$_{71}$ (see Table 
      \ref{elements}) are suggestive of a NEO that moves co-orbitally with the Earth. The source of the Heliocentric Keplerian 
      osculating orbital elements and uncertainties in Table \ref{elements} is the JPL Small-Body Database.\footnote{\url{http://ssd.jpl.nasa.gov/sbdb.cgi}} 
      Its current orbit is reliable enough (see discussion above) to assess its short-term dynamical evolution as it is based on 
      34 observations for a data-arc span of 21 d. The quality of the orbit of this object is at present lower than that of 
      2014~EK$_{24}$ or 2014~UR and is on par with that of 2014~QD$_{364}$ (see Table \ref{elements}). In addition, it is similar 
      (34 observations spanning 21 d versus 18 observations spanning 24 d) to that of the orbital solution of 2013~LX$_{28}$, 
      another very stable Earth co-orbital, when it was recognised as a quasi-satellite of our planet by Connors (2014). 

      With a value of the semi-major axis $a$ = 0.9885 AU, very close to that of our planet (0.9992 AU), this Aten asteroid is a 
      NEO moving in a very Earth-like orbit with low-eccentricity, $e$ = 0.09, and little inclination, $i$ = 4\fdg9. With such an 
      orbit, close encounters are only possible with the Earth--Moon system, and they take place with a cadence equal to its 
      synodic period (see above). One of the reasons for its inclusion here is to encourage further observations of this 
      interesting and unusual (at least dynamically) minor body that is characterised by very sparse favourable visibility 
      windows; the last one occurred around 1959. This object will experience close encounters with our planet under 0.5 AU in 
      June 2015 (0.18 AU), 2016 (0.28 AU), 2017 (0.39 AU), and 2018 (0.50 AU), but after this last relatively favourable flyby, 
      the next one under 0.5 AU will take place in 2061 February (0.44 AU). 
%
%---------------------------------------------------------------------------------------------------------------------------------
%
      \onlfig{
      \begin{figure*}
        \centering
         \includegraphics[width=\linewidth]{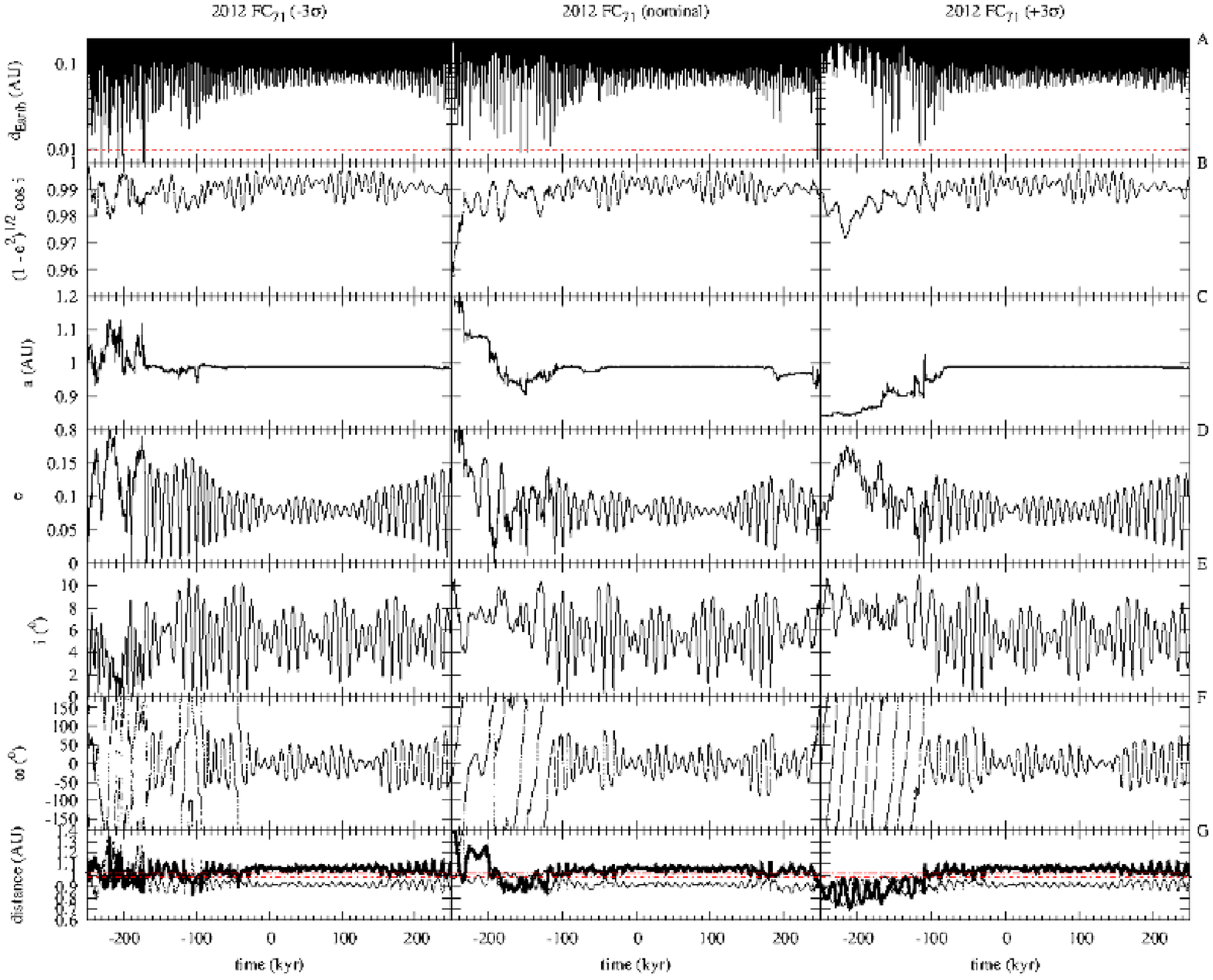}
         \caption{Comparative short-term dynamical evolution of various parameters for the nominal orbit of 2012~FC$_{71}$ as 
                  presented in Table \ref{elements} (central panels) and two representative examples of orbits that are the most 
                  different from the nominal one ($\pm3\sigma$ deviations, see  text for details). The distance from the Earth 
                  (panel A); the value of the Hill sphere radius of the Earth, 0.0098 AU, is displayed (red line). The parameter 
                  $\sqrt{1 - e^2} \cos i$ (B). The orbital elements $a$ (C), $e$ (D), $i$ (E) and 
                  $\omega$ (F). The distances to the descending (thick line) and ascending nodes (dotted line) appear in 
                  panel G. Earth's aphelion and perihelion distances are also shown (red lines).
                 }
         \label{controlfc71}
      \end{figure*}
      }
%
%---------------------------------------------------------------------------------------------------------------------------------
%

\begin{figure*}
        \centering
         \includegraphics[width=\linewidth]{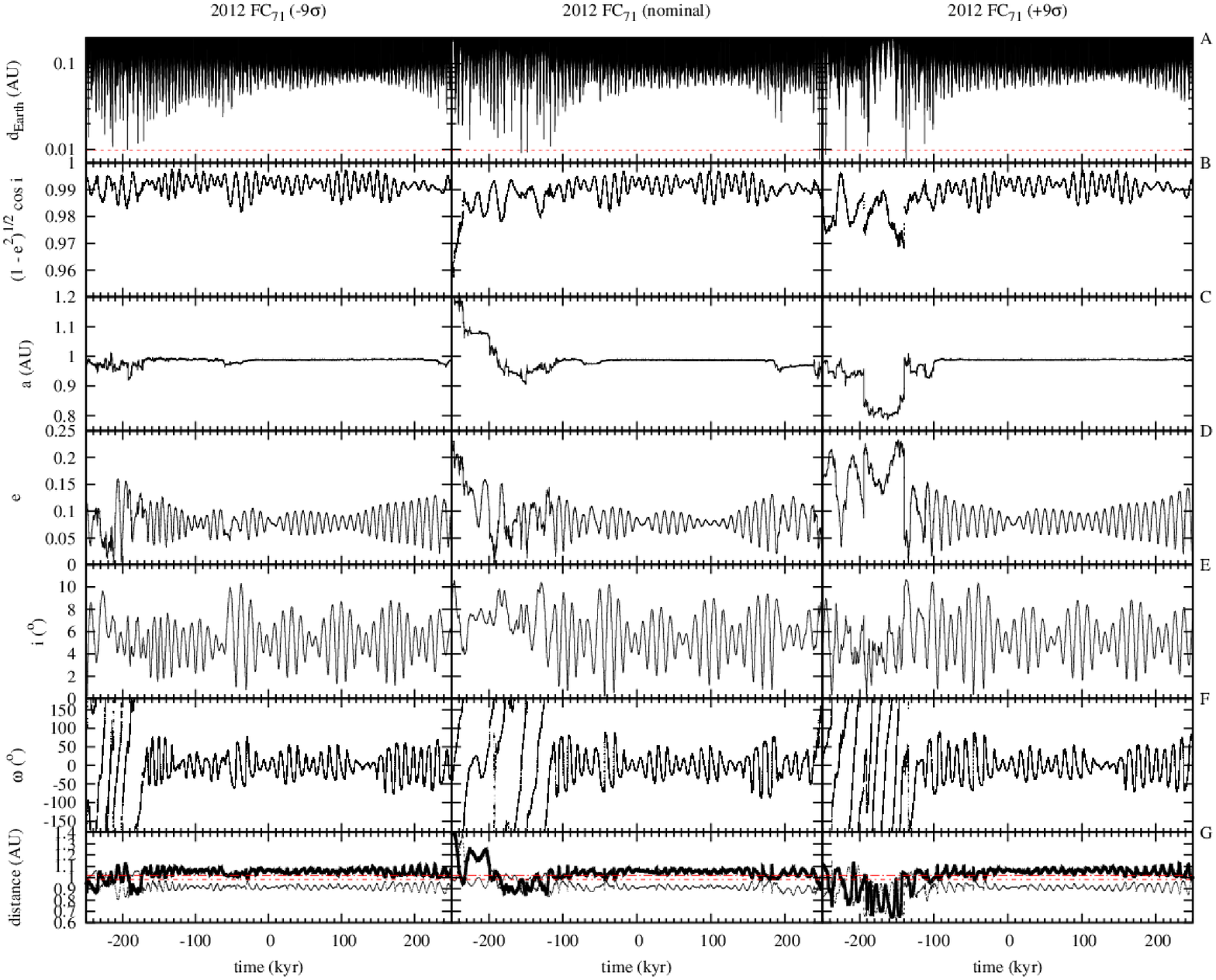}
         \caption{Comparative short-term dynamical evolution of various parameters for the nominal orbit of 2012~FC$_{71}$ as 
                  presented in Table \ref{elements} (central panels) and two representative examples of orbits that are the most 
                  different from the nominal one ($\pm9\sigma$ deviations, see text for details). The distance from the Earth 
                  (panel A); the value of the Hill sphere radius of the Earth, 0.0098 AU, is shown (red line). The parameter 
                  $\sqrt{1 - e^2} \cos i$ (B). The orbital elements $a$ (C), $e$ (D), $i$ (E), and $\omega$ (F). The distances to 
                  the descending (thick line) and ascending nodes (dotted line) appear in panel G. Earth's aphelion and perihelion 
                  distances are also shown (red lines).
                 }
         \label{controlfc712}
      \end{figure*}

      In order to study the possible co-orbital nature of 2012~FC$_{71}$ with the Earth, we have performed $N$-body calculations 
      in both directions of time for 250 kyr using the physical model described above. Regular co-orbitals are characterised by 
      the libration of the relative mean longitude, $\lambda_{\rm r}$, or the difference between the mean longitudes of the object 
      and its host planet. The mean longitude of an object is given by $\lambda$ = $M$ + $\Omega$ + $\omega$, where $M$ is the 
      mean anomaly, $\Omega$ is the longitude of ascending node, and $\omega$ is the argument of perihelion  (see e.g. Murray \& 
      Dermott 1999). Asteroid 2012~FC$_{71}$ is not a co-orbital of the Earth in the classical sense (see e.g. Morais \& 
      Morbidelli 2002) because its $\lambda_{\rm r}$ does not currently librate but circulates (not shown in the figures); 
      however, its $\omega$ does librate around 0\degr\ which means that it is submitted to a secular resonance, see panel F in 
      Figs. \ref{controlfc71} (available electronically only) and \ref{controlfc712}, the Kozai resonance (Kozai 1962). There is 
      no libration of $\lambda_{\rm r}$ because $\dot{\omega}_{\rm r}$ librates about 0\degr. In a Kozai resonance, the apse and 
      the node are in resonance with one another (Kozai 1962). Because of the Kozai resonance, both eccentricity and inclination 
      oscillate with the same frequency but out of phase; when the value of the eccentricity reaches its maximum the value of the 
      inclination is the lowest and vice versa ($\sqrt{1 - e^2} \cos i \sim$ constant, panel B in Figs. \ref{controlfc71} and 
      \ref{controlfc712}). The values of the eccentricity and inclination are coupled, and the value of the semi-major axis 
      remains nearly constant (see panels C, D, and E in Figs. \ref{controlfc71} and \ref{controlfc712}). At a proper average 
      inclination of 5\fdg5, this object was --when discovered-- the coldest (dynamically speaking) known Kozai resonator (de 
      la Fuente Marcos \& de la Fuente Marcos 2013). Asteroids 2014~EK$_{24}$ and 2014~QD$_{364}$ are even colder Kozai librators,
      but are less stable. 
%
%---------------------------------------------------------------------------------------------------------------------------------
%
%
%---------------------------------------------------------------------------------------------------------------------------------
%

      Figure \ref{controlfc71} shows the short-term dynamical evolution of three illustrative orbits, the nominal one (central 
      panels) and those of two representative orbits that are the most different from the nominal one. In Figs. \ref{controlfc71} 
      to \ref{controlur} when an orbit is labelled `$\pm{n}\sigma$', where $n$ is an integer, it has been obtained by adding (+) 
      or subtracting ($-$) $n$-times the uncertainty from the orbital parameters (the six elements) in Table \ref{elements}. All 
      the control orbits exhibit consistent behaviour within a few hundred thousand years of $t = 0$. This object can only 
      experience close encounters with the Earth--Moon system, see panel G in Fig. \ref{controlfc71}. Close encounters place the 
      object inside the Kozai resonance, see panel A in Fig. \ref{controlfc71}. To further support the idea that the relatively 
      low quality of the present orbit of this particular object is not an obstacle to obtaining a clear and reliable picture of 
      its past and future orbital evolution, Fig. \ref{controlfc712} shows additional control orbits where the orbital elements 
      have been further modified at the $\pm9\sigma$ level. The short-term dynamical evolution of these orbits is still consistent 
      with that in Fig. \ref{controlfc71}. Although based on a relatively short data-arc (21 d), the orbital evolution of 
      2012~FC$_{71}$ is remarkably stable. The nodes of 2012~FC$_{71}$ remain far from our planet for very long periods of time 
      (see panel G in Figs. \ref{controlfc71} and \ref{controlfc712}). The present quality of the orbit is obviously not an issue 
      regarding the study of the dynamical evolution of this object. The duration of its current co-orbital state is at least 330 
      kyr, but episodes lasting nearly 500 kyr are feasible. Longer calculations, spanning 3 Myr, show that recurrent episodes are 
      also possible.

   \section{Asteroid 2014~EK$_{24}$, an Apollo Kozai librator}
      Asteroid 2014~EK$_{24}$ was discovered on 2014 March 10 by S.~M.~Larson observing for the Catalina Sky Survey with the 
      0.68 m Schmidt telescope (Larson et al. 2014). It is larger than 2012~FC$_{71}$ at $H$ = 23.2 or a diameter in the range 
      60--150 m if an albedo of 0.20--0.04 is assumed. It is the largest of the objects studied here. As in the previous case, the 
      orbital elements of 2014~EK$_{24}$ (see Table \ref{elements}) resemble those of a co-orbital NEO. Its orbit is well defined 
      as it is based on 267 observations with a data-arc span of 390 d. The value of its semi-major axis is also very close to 
      that of our planet, $a$ = 1.0043 AU. This Apollo asteroid is a NEO that also follows a very Earth-like orbit with $e$ = 
      0.07 and $i$ = 4\fdg7. As in the case of 2012~FC$_{71}$, its path is only directly perturbed by the Earth--Moon system 
      during relatively distant close encounters (the minimum distance is currently $>$0.03 AU). Asteroid 2014~EK$_{24}$ is 
      included on NASA's list of potential human mission targets (the NHATS list), which means that improving our understanding of 
      the dynamics of this object is particularly important. Based on observations made with the IAC80 and ESA-OGS telescopes
      operated in Tenerife by the Instituto de Astrof\'isica de Canarias (IAC) in the Spanish Observatorio del Teide, Radu Cornea 
      and Ovidiu Vaduvescu (pers. comm.) from the EURONEAR network (Vaduvescu et al. 2008, 2015) have reported a rotation period 
      of 6 minutes. Its very fast rotation is also characterised by a large amplitude of 0.79 mag. Elongated asteroids have large 
      lightcurve amplitudes (see e.g. Kaasalainen \& Torppa 2001).

\begin{figure*}
        \centering
         \includegraphics[width=\linewidth]{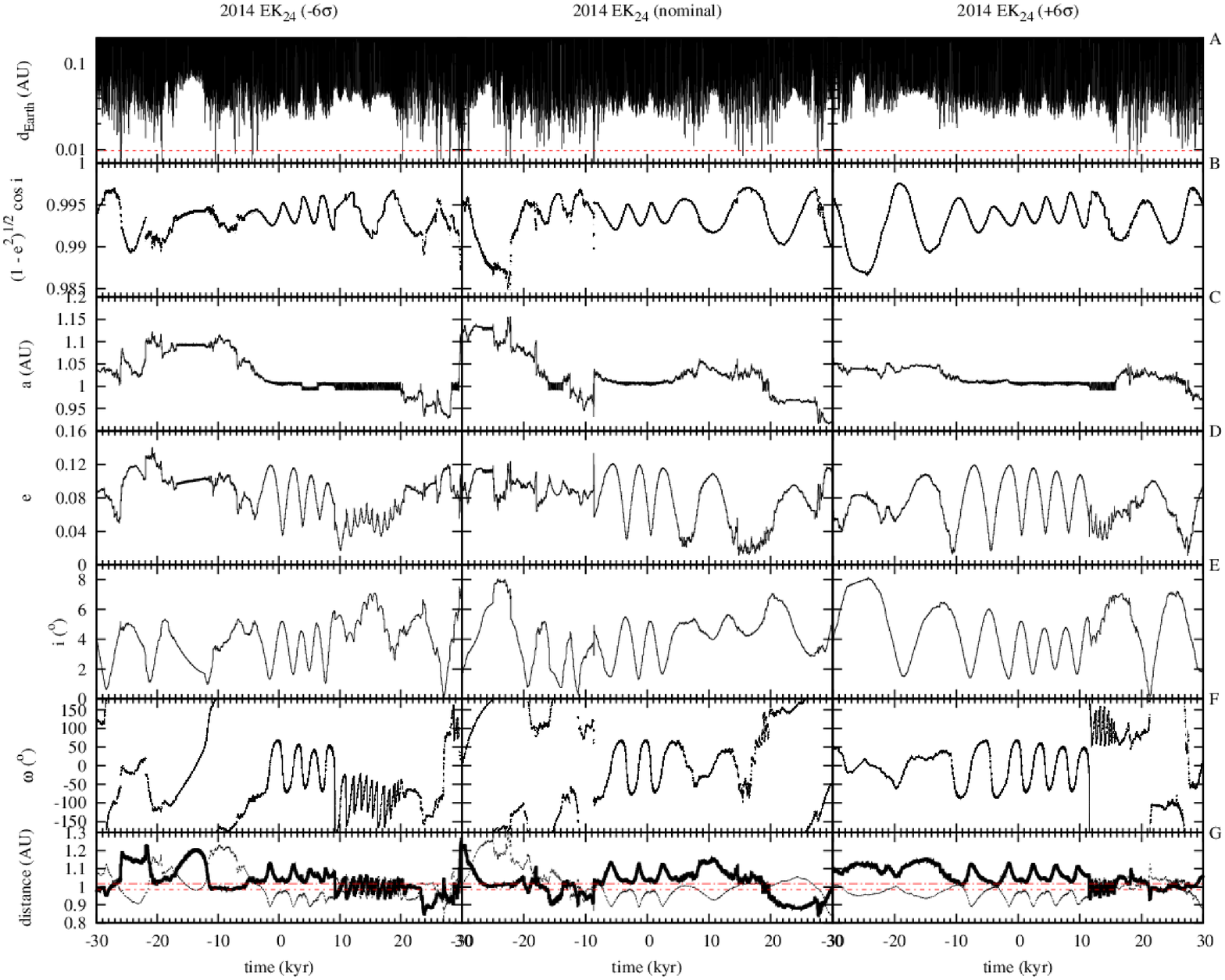}
         \caption{Same as Fig. \ref{controlfc71} but for 2014~EK$_{24}$ and $\pm6\sigma$ (see text for details).
                 }
         \label{controlek24}
      \end{figure*}

      Figure \ref{controlek24} shows the short-term dynamical evolution of this object under the same conditions used for 
      2012~FC$_{71}$ except that only the time interval ($-$30, 30) kyr is given because the object is considerably less stable. 
      In Fig. \ref{controlek24} we use three illustrative orbits, the nominal one (central panels) and those of two representative 
      orbits that are the most different from the nominal one. A smaller value of the dispersion has been used because the orbit 
      of this object is considerably more robust, statistically speaking. All the control orbits exhibit coherent behaviour within 
      a few thousand years of $t = 0$. As in the previous case and for the same reason, its $\lambda_{\rm r}$ does not currently 
      librate (not shown), but its orbital period (1.01 yr) matches that of the Earth and its $\omega$ librates around 0\degr (see 
      panel F in Fig. \ref{controlek24}), the signposts of a co-orbital passing orbit with a small Jacobi constant. This 
      assessment is statistically robust according to the current observational uncertainties or at a confidence level $>$ 99.99\% 
      (6$\sigma$). Brief ($\sim$40--200 yr) co-orbital episodes of the horseshoe and quasi-satellite type are also observed. 
      During these episodes $\lambda_{\rm r}$ librates and $\dot{\omega}_{\rm r}$ is positive (horseshoe) or negative 
      (quasi-satellite). The value of the parameter $\sqrt{1 - e^2} \cos i$ (see panel B in Fig. \ref{controlek24}) remains 
      approximately constant for most of the evolution shown.

      This minor body started following a co-orbital passing orbit nearly 4.5 kyr ago and it will continue in this state for at 
      least 5.5 more kyr. The duration of the entire episode is in the range 15 to 40 kyr with shorter episodes being more likely. 
      Asteroid 2014~EK$_{24}$ is an even colder (dynamically speaking) Kozai resonator than 2012~FC$_{71}$, with a proper average 
      inclination of just 3\fdg3 (see panel E in Fig. \ref{controlek24}), which means that the encounters with the Earth--Moon 
      system are closer and take place more often than in the case of 2012~FC$_{71}$ (compare panel A in Figs. \ref{controlfc71} 
      or \ref{controlfc712} and \ref{controlek24}). This explains why 2014~EK$_{24}$ is overall less dynamically stable than 
      2012~FC$_{71}$. Close encounters are responsible for both injection into and ejection from the Kozai state.  
%
%---------------------------------------------------------------------------------------------------------------------------------
%
%
%---------------------------------------------------------------------------------------------------------------------------------
%

   \section{Asteroid 2014~QD$_{364}$, a small Aten Kozai librator}
      Asteroid 2014~QD$_{364}$ was discovered on 2014 August 30 by the Catalina Sky Survey (Kowalski et al. 2014). Its orbit is 
      reasonably well known and is based on 29 observations with a data-arc span of 17 d similar to that of 2012~FC$_{71}$, 
      although the orbit is significantly less stable. In principle, it is the smallest of the four objects studied here at $H$ = 
      27.2 or a diameter in the range 10--24 m if an albedo of 0.20--0.04 is assumed. As in the previous two cases, the orbital 
      parameters of this object (see Table \ref{elements}) are comparable to those of a co-orbital NEO. The value of its 
      semi-major axis is very close to that of our planet, $a$ = 0.9891 AU. This Aten asteroid is a NEO that follows a very 
      Earth-like orbit with $e$ = 0.04 and $i$ = 4\fdg0; its path is at present only directly perturbed by the Earth--Moon system 
      during relatively distant close encounters (currently the minimum distance is $>$0.01 AU), well separated in time (nearly 60 
      yr). 
      
\begin{figure*}
        \centering
         \includegraphics[width=\linewidth]{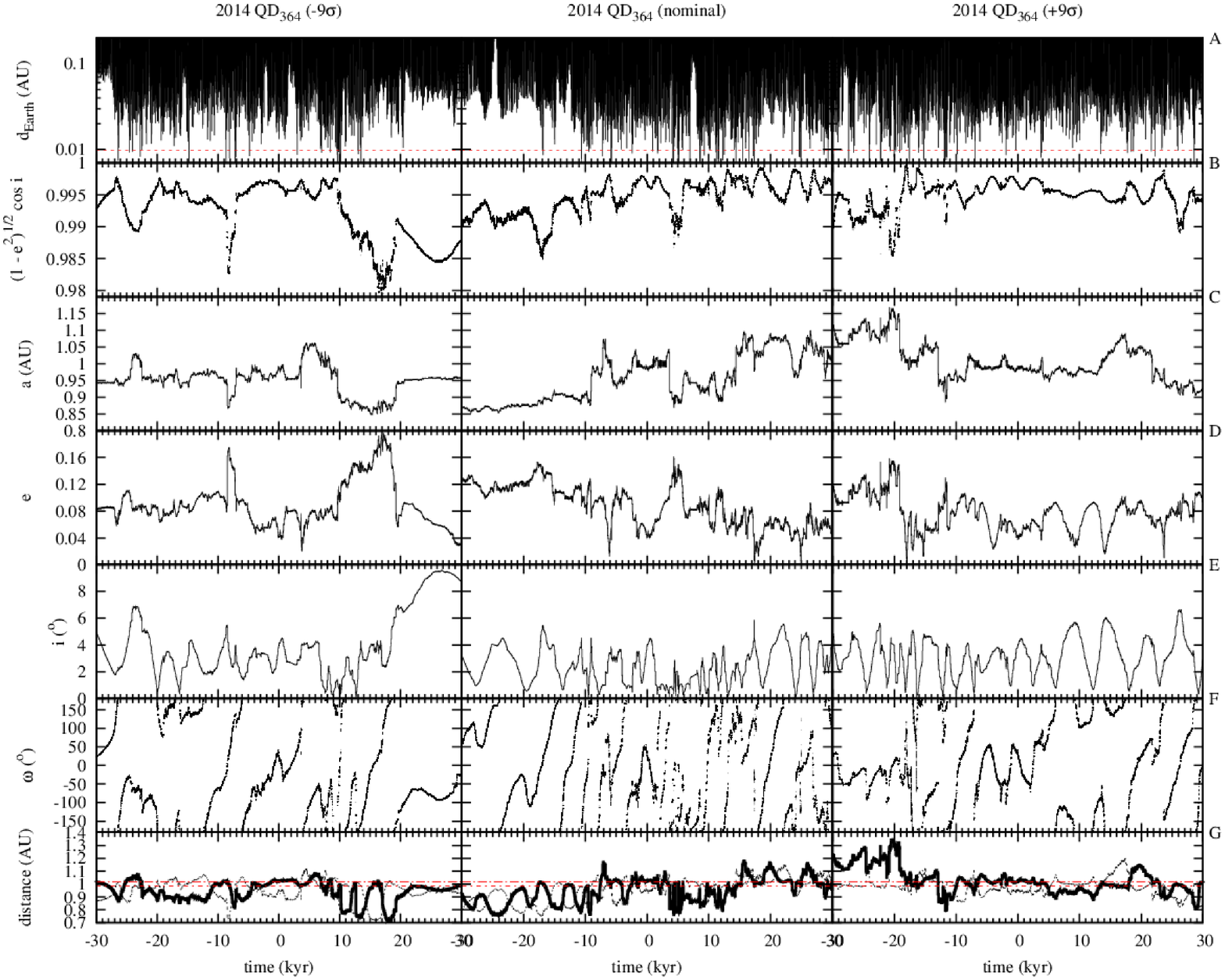}
         \caption{Same as Fig. \ref{controlfc71} but for 2014~QD$_{364}$ and $\pm9\sigma$ (see text for details).
                 }
         \label{controlqd364}
      \end{figure*}

      Figure \ref{controlqd364} shows the short-term dynamical evolution of this object under the same conditions used for the 
      previous two. As in the case of 2014~EK$_{24}$, only the time interval ($-$30, 30) kyr is shown because the object is 
      not very stable; in fact, it is perhaps the most unstable of the four minor bodies studied here. In Fig. \ref{controlqd364} 
      (following the same reasoning applied to 2012~FC$_{71}$), we display three illustrative orbits, the nominal one (central 
      panels) and those of two representative orbits that are the most different from the nominal one. All the control orbits 
      exhibit consistent behaviour within a few thousand years of $t = 0$.  
%
%---------------------------------------------------------------------------------------------------------------------------------
%
%
%---------------------------------------------------------------------------------------------------------------------------------
%

      Asteroid 2014~QD$_{364}$ appears to be precariously perched on the threshold of instability. The descending node is 
      dangerously close to Earth's aphelion (see panel G in Fig. \ref{controlqd364}). Very similar control orbits evolve rather 
      differently after just a few thousand years. It began following a co-orbital passing orbit nearly 2 kyr ago and it will 
      continue doing so for at least the next 1.5 kyr. As in the previous two cases, its argument of perihelion currently librates 
      around 0\degr. Close encounters are responsible for both injection into and ejection from the Kozai state that lasts at 
      least 4 kyr and as long as 10 kyr, the shortest episodes being more common. Kozai states characterised by the libration of 
      the argument of perihelion around 90\degr, $-90\degr$, or 180\degr\ for a few thousand years are possible. Brief ($\sim$40 
      yr to 1.5 kyr) co-orbital episodes of the horseshoe and quasi-satellite type are also observed. The horseshoe episodes last 
      longer.

   \section{Asteroid 2014~UR, yet another small Aten Kozai librator}
      Asteroid 2014~UR was discovered on 2014 October 17, also by the Catalina Sky Survey (Hill et al. 2014). It is smaller than 
      2014~EK$_{24}$ and 2012~FC$_{71}$ but larger than 2014~QD$_{364}$ at $H$ = 26.6 or a diameter in the range 14--32 m if an 
      albedo of 0.20--0.04 is assumed. As in the previous cases, the orbital parameters of 2014~UR (see Table \ref{elements}) are 
      similar to those of a co-orbital NEO. Its orbital solution is very robust and it is based on 146 observations, including 2 
      radar observations, with a data-arc span of 14 d. Its orbit uncertainty is similar to that of 2014~EK$_{24}$. The value of 
      its semi-major axis ($a$ = 0.9991 AU) is very close to that of our planet. This Aten asteroid is a NEO that also follows an 
      Earth-like orbit with $e$ = 0.01 and $i$ = 8\fdg2. Like the previous three, its path is at the moment only directly 
      perturbed by the Earth--Moon system during close encounters (the minimum distance is currently $>$0.008 AU). 

\begin{figure*}
        \centering
         \includegraphics[width=\linewidth]{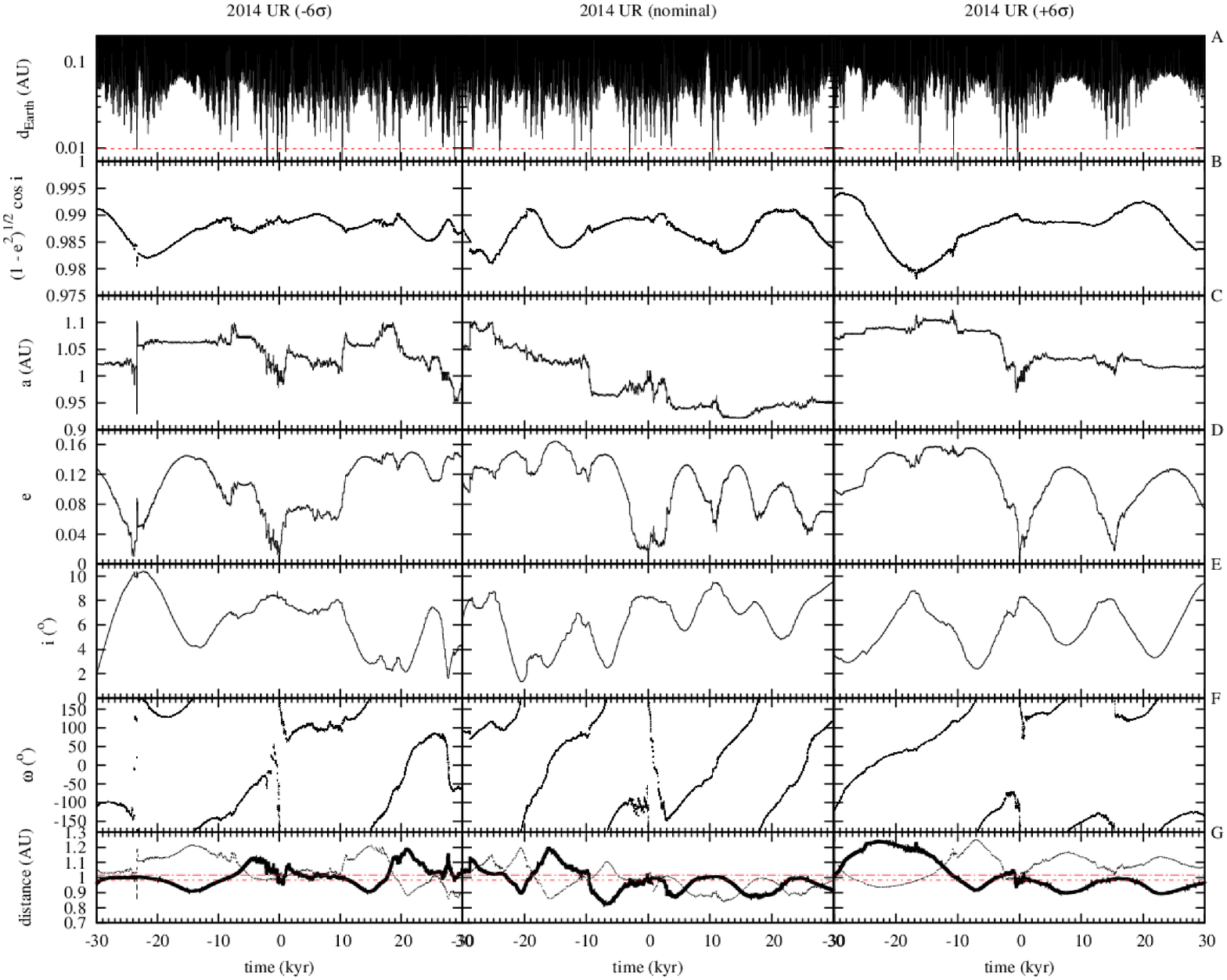}
         \caption{Same as Fig. \ref{controlfc71} but for 2014~UR and $\pm6\sigma$ (see  text for details).
                 }
         \label{controlur}
      \end{figure*}

      Figure \ref{controlur} shows the short-term dynamical evolution of this minor body during the time interval ($-30$, 30) kyr. 
      It presents the past and future behaviour of the nominal solution as well as those of two additional representative control 
      orbits where the orbital elements have been modified at the $\pm6\sigma$ level (see above). Its overall past and future 
      dynamical evolution is quite uncertain and it may be as unstable and chaotic as that of 2014~QD$_{364}$. Some of the control 
      orbits show long Kozai episodes in which the argument of perihelion librates around 180\degr for many thousands of years 
      (see right-hand panel F in Fig. \ref{controlur}). The close encounters with the Earth--Moon system experienced by 2014~UR 
      are stronger than those of the previous three objects and that explains why it is so unstable.
%
%---------------------------------------------------------------------------------------------------------------------------------
%
%
%---------------------------------------------------------------------------------------------------------------------------------
%

   \section{Relative eccentricity-relative argument of perihelion}
\begin{figure}
        \centering
         \includegraphics[width=\linewidth]{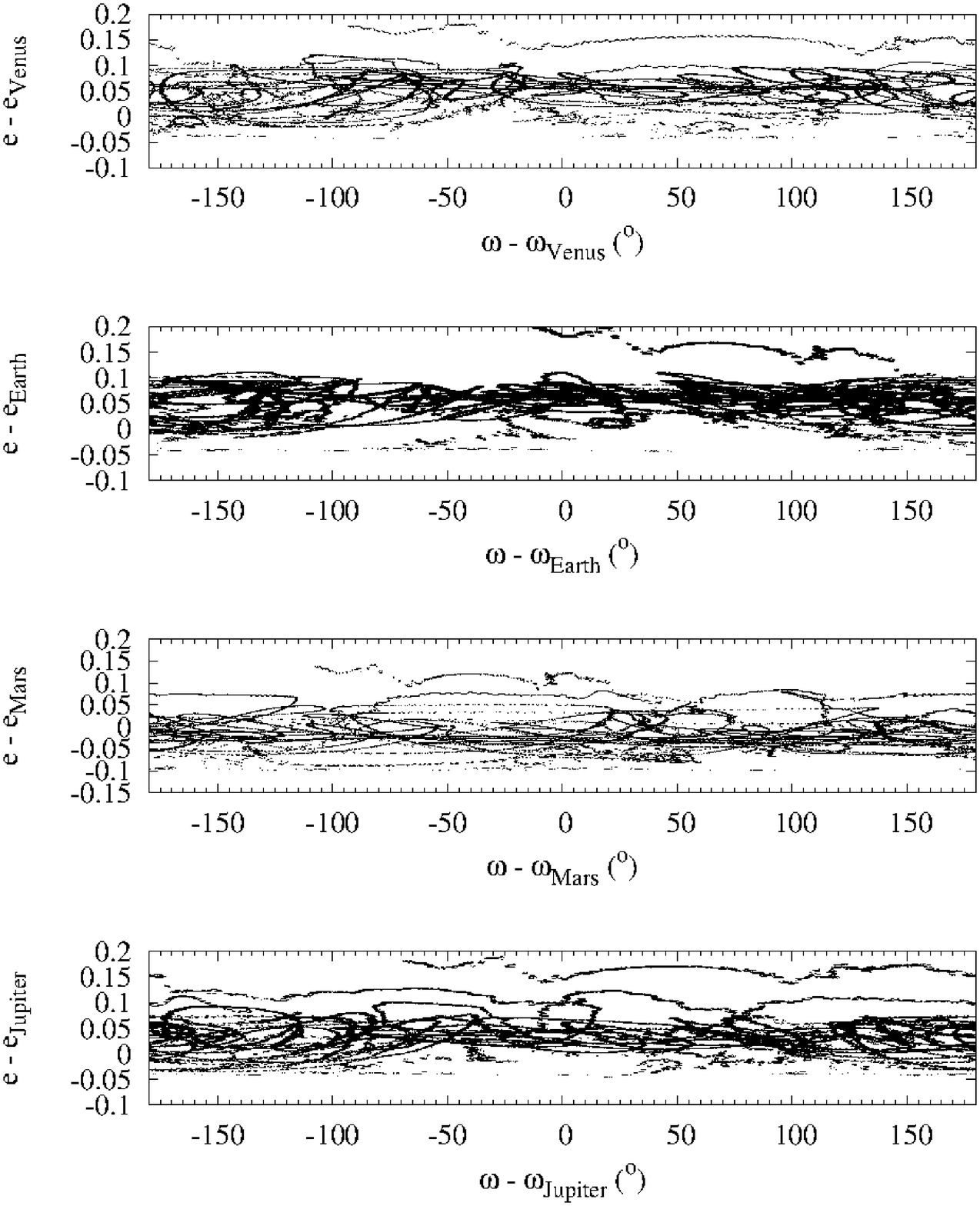}
         \caption{The $e_{\rm r} \omega_{\rm r}$-portrait relative to Venus, the Earth, Mars, and Jupiter for 2012~FC$_{71}$.
                 }
         \label{fEWfc71M}
      \end{figure}

      The secular evolution of co-orbital objects is viewed naturally in the $e_{\rm r} \omega_{\rm r}$-plane, where $e_{\rm r} = 
      e - e_{\rm p}$ and $\omega_{\rm r} = \omega - \omega_{\rm p}$; $e_{\rm p}$ and $\omega_{\rm p}$ are, respectively, the 
      eccentricity and argument of perihelion of a given planet. In Namouni (1999), the analysis is carried out within the 
      framework of the restricted elliptic three-body problem; his fig. 19 shows a selection of orbits in the $e_{\rm r} 
      \omega_{\rm r}$-plane and they all look very regular. In sharp contrast, our $e_{\rm r} \omega_{\rm r}$ maps for 
      2012~FC$_{71}$, 2014~EK$_{24}$, 2014~QD$_{364}$, and 2014~UR relative to the Earth (second from top panels in Figs. 
      \ref{fEWfc71M}, \ref{fEWek24M}, \ref{fEWqd364M}, and \ref{fEWurM}) display a rather convoluted evolution, in particular 
      those of 2012~FC$_{71}$ and 2014~QD$_{364}$. We cannot conclude that the Kozai dynamics followed by these objects are the 
      result of the dominant secular perturbation of Venus, Earth, Mars, or Jupiter. This superposition creates very rich dynamics 
      instead. 

      In addition to being trapped in a near 1:1 mean motion resonance with our planet, these asteroids orbit the Sun in a near 
      13:8 mean motion resonance with Venus, so this planet completes 13 orbits around the Sun in the same amount of time that the 
      asteroids complete 8; they also move in a near 2:1 mean motion resonance with Mars, so Mars completes one orbit while the 
      minor bodies do two and a 12:1 mean motion resonance with Jupiter, completing 12 orbits around the Sun approximately in the 
      same amount of time Jupiter completes only one. These near resonances induce the dramatic changes observed in the $e_{\rm r} 
      \omega_{\rm r}$-portraits. The irregular evolution displayed in Figs. \ref{fEWfc71M}, \ref{fEWek24M}, \ref{fEWqd364M}, and 
      \ref{fEWurM} shows that the dynamics of these objects cannot be properly described within the framework of the restricted 
      elliptic three-body problem alone, at least not during the entire time-span studied here.
%
%---------------------------------------------------------------------------------------------------------------------------------
%
%
%---------------------------------------------------------------------------------------------------------------------------------
%
%
%---------------------------------------------------------------------------------------------------------------------------------
%
      \onlfig{
      \begin{figure}
        \centering
         \includegraphics[width=\linewidth]{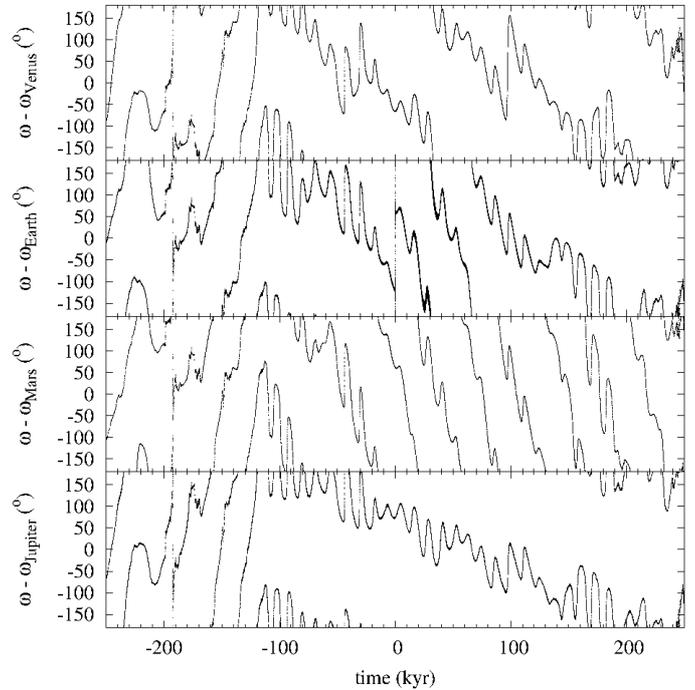}
         \caption{Time evolution of $\omega_{\rm r}$ relative to Venus, the Earth, Mars, and Jupiter for 2012~FC$_{71}$.
                 }
         \label{fEWfc71T}
      \end{figure}
      }
%
%---------------------------------------------------------------------------------------------------------------------------------
%
%
%---------------------------------------------------------------------------------------------------------------------------------
%
      \begin{figure}
        \centering
         \includegraphics[width=\linewidth]{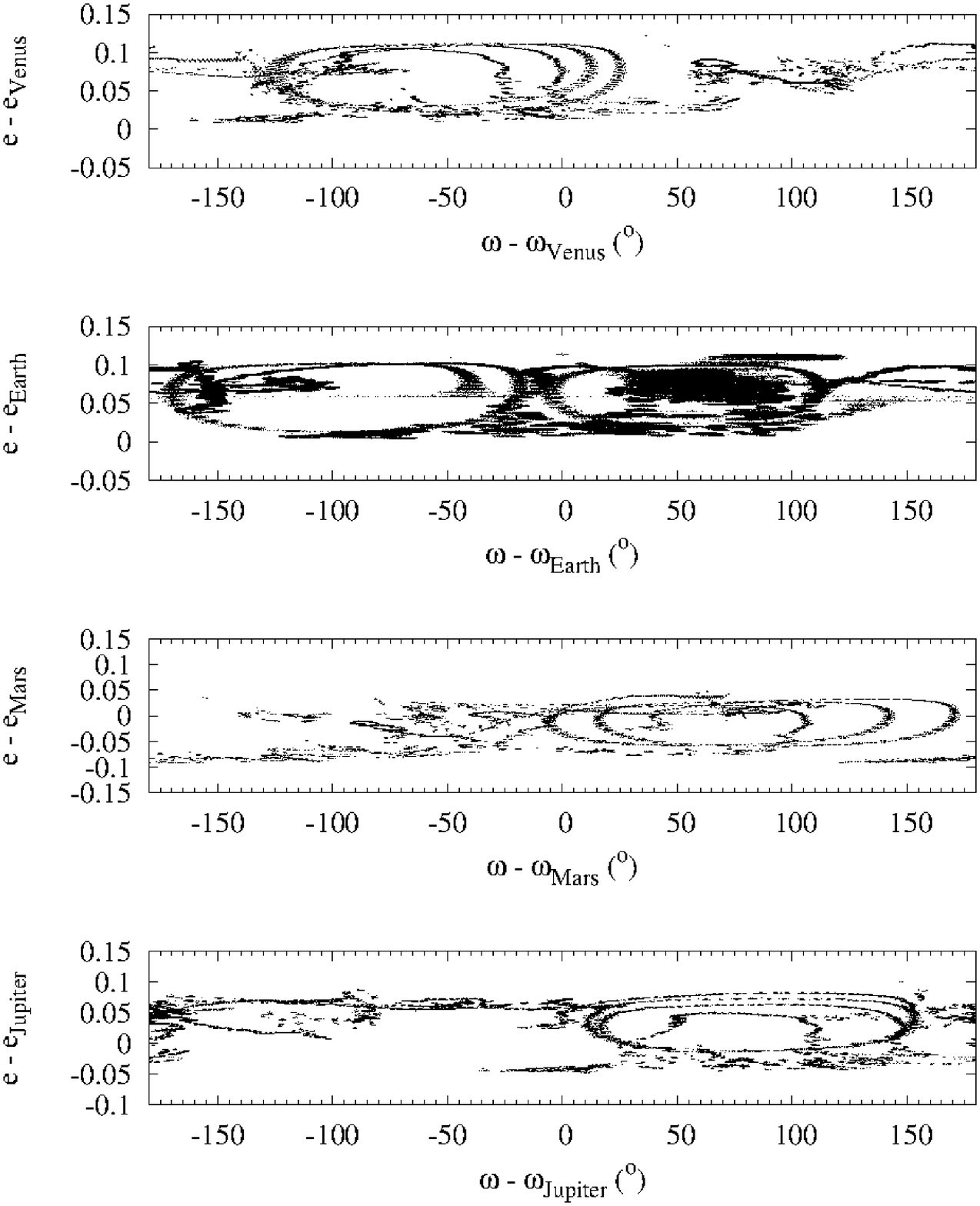}
         \caption{Same as Fig. \ref{fEWfc71M} but for 2014~EK$_{24}$.
                 }
         \label{fEWek24M}
      \end{figure}
%
%---------------------------------------------------------------------------------------------------------------------------------
%
%
%---------------------------------------------------------------------------------------------------------------------------------
%
      \onlfig{
      \begin{figure}
        \centering
         \includegraphics[width=\linewidth]{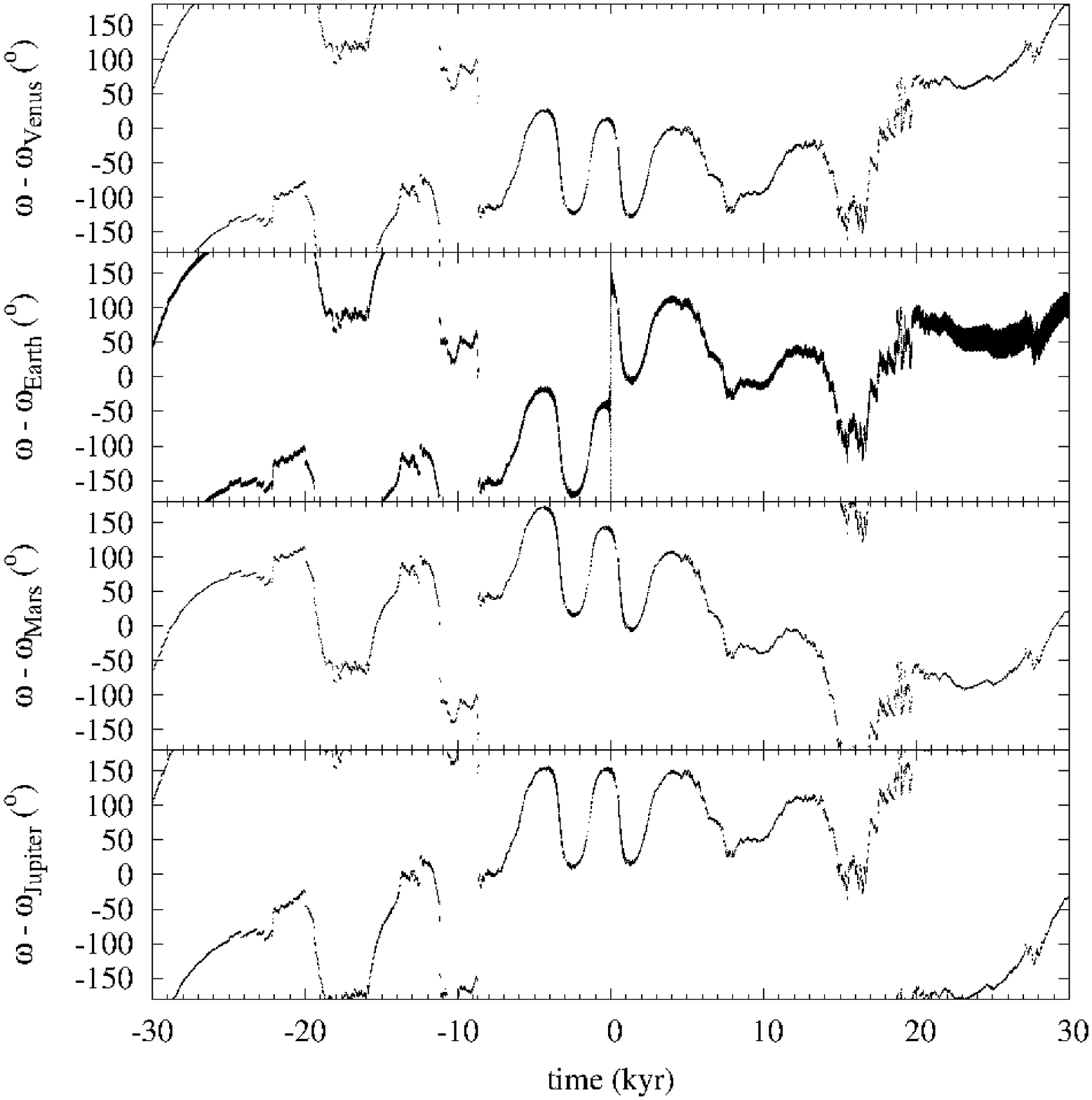}
         \caption{Same as Fig. \ref{fEWfc71T} but for 2014~EK$_{24}$.
                 }
         \label{fEWek24T}
      \end{figure}
      }
%
%---------------------------------------------------------------------------------------------------------------------------------
%
%
%---------------------------------------------------------------------------------------------------------------------------------
%
      \begin{figure}
        \centering
         \includegraphics[width=\linewidth]{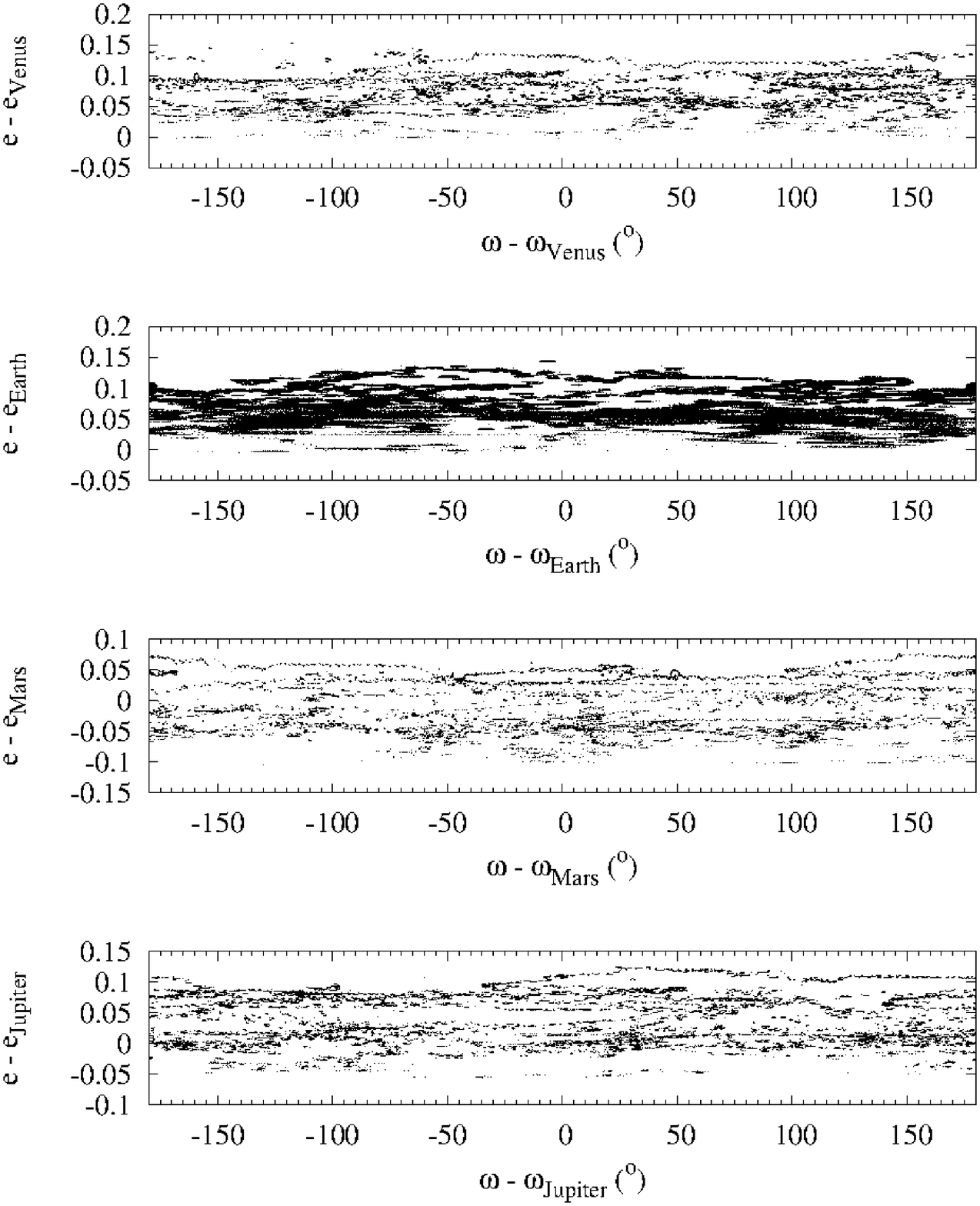}
         \caption{Same as Fig. \ref{fEWfc71M} but for 2014~QD$_{364}$.
                 }
         \label{fEWqd364M}
      \end{figure}
%
%---------------------------------------------------------------------------------------------------------------------------------
%
%
%---------------------------------------------------------------------------------------------------------------------------------
%
      \onlfig{
      \begin{figure}
        \centering
         \includegraphics[width=\linewidth]{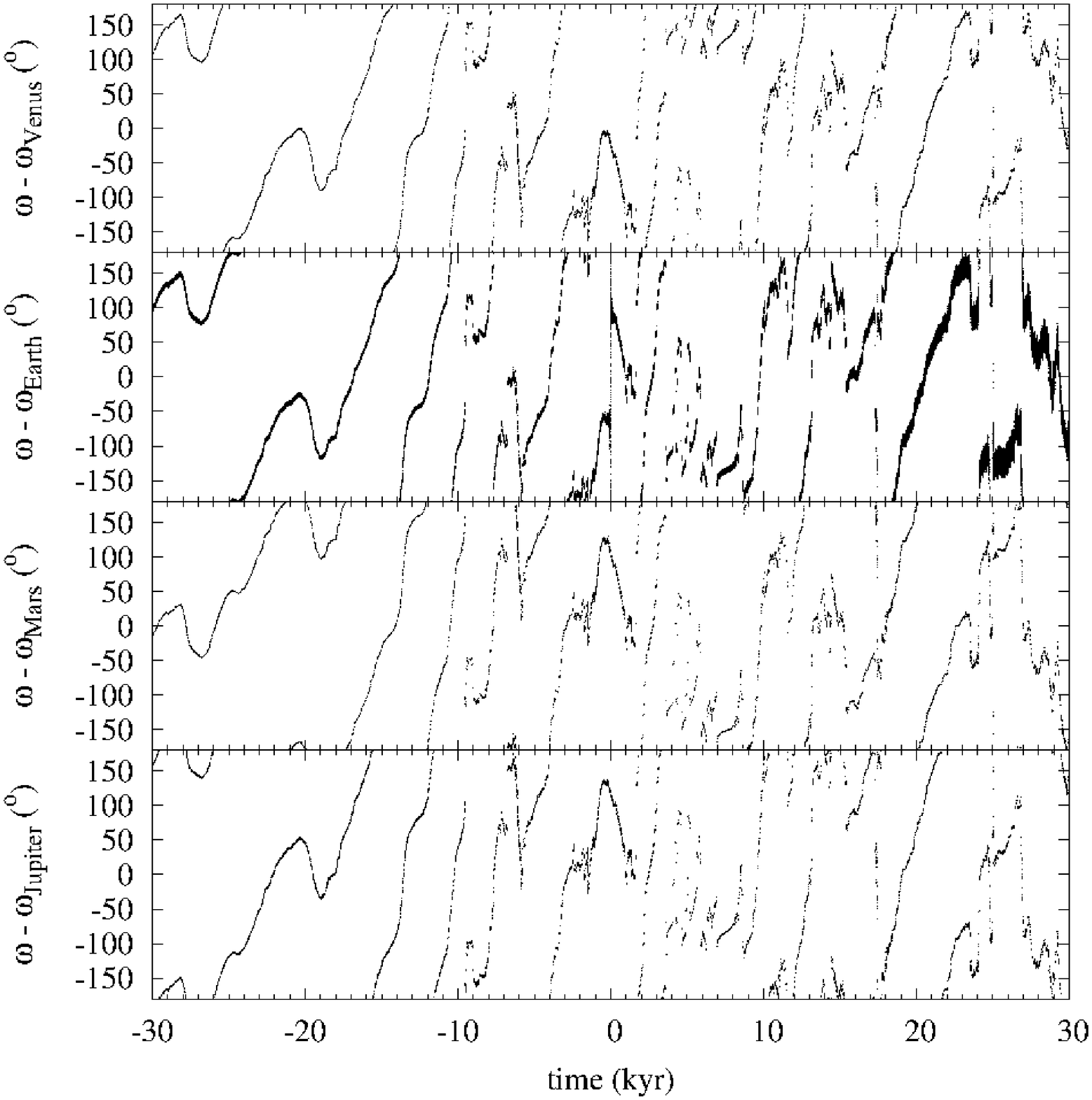}
         \caption{Same as Fig. \ref{fEWfc71T} but for 2014~QD$_{364}$.
                 }
         \label{fEWqd364T}
      \end{figure}
      }
%
%---------------------------------------------------------------------------------------------------------------------------------
%
%
%---------------------------------------------------------------------------------------------------------------------------------
%
      \begin{figure}
        \centering
         \includegraphics[width=\linewidth]{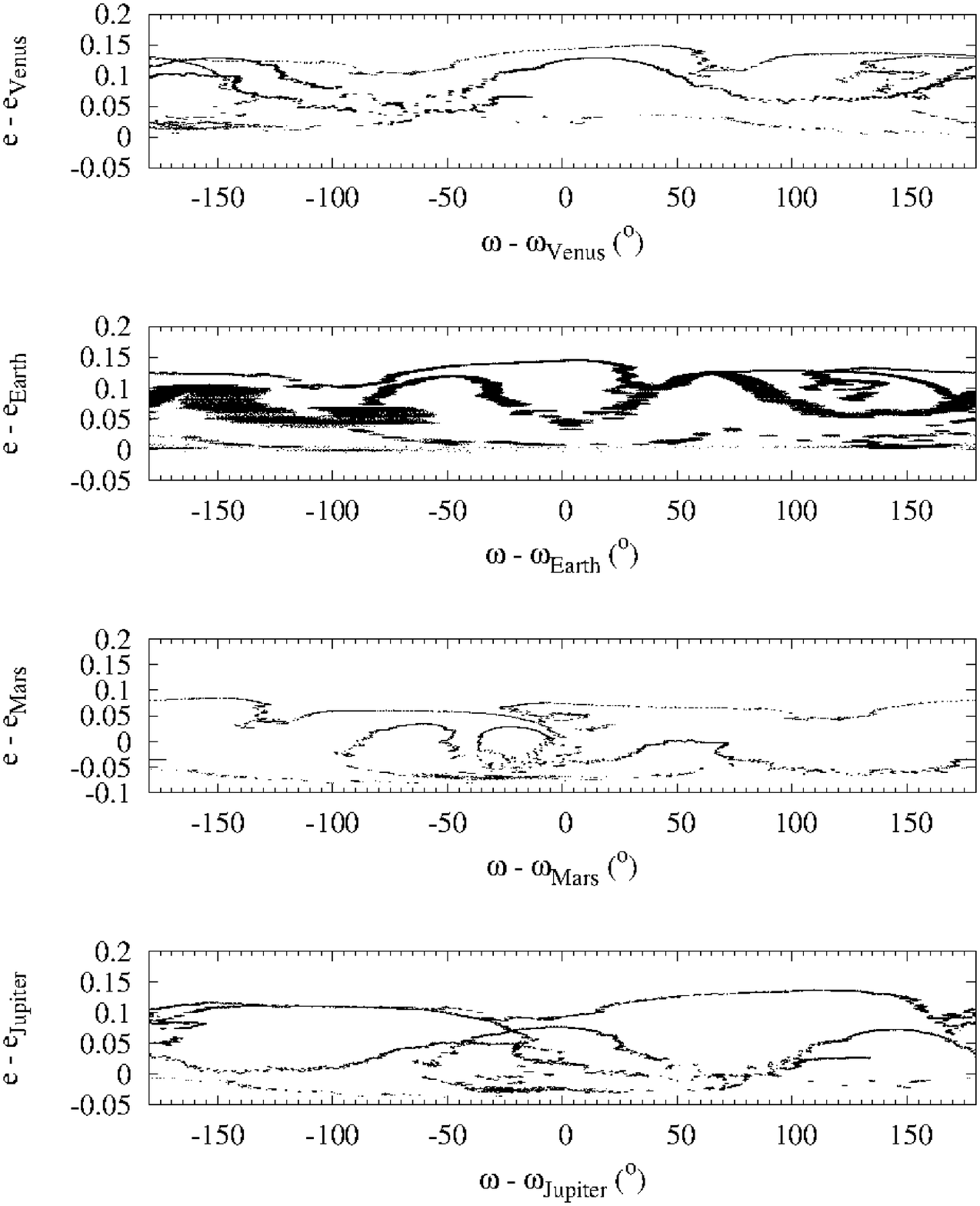}
         \caption{Same as Fig. \ref{fEWfc71M} but for 2014~UR.
                 }
         \label{fEWurM}
      \end{figure}
%
%---------------------------------------------------------------------------------------------------------------------------------
%
%
%---------------------------------------------------------------------------------------------------------------------------------
%
      \onlfig{
      \begin{figure}
        \centering
         \includegraphics[width=\linewidth]{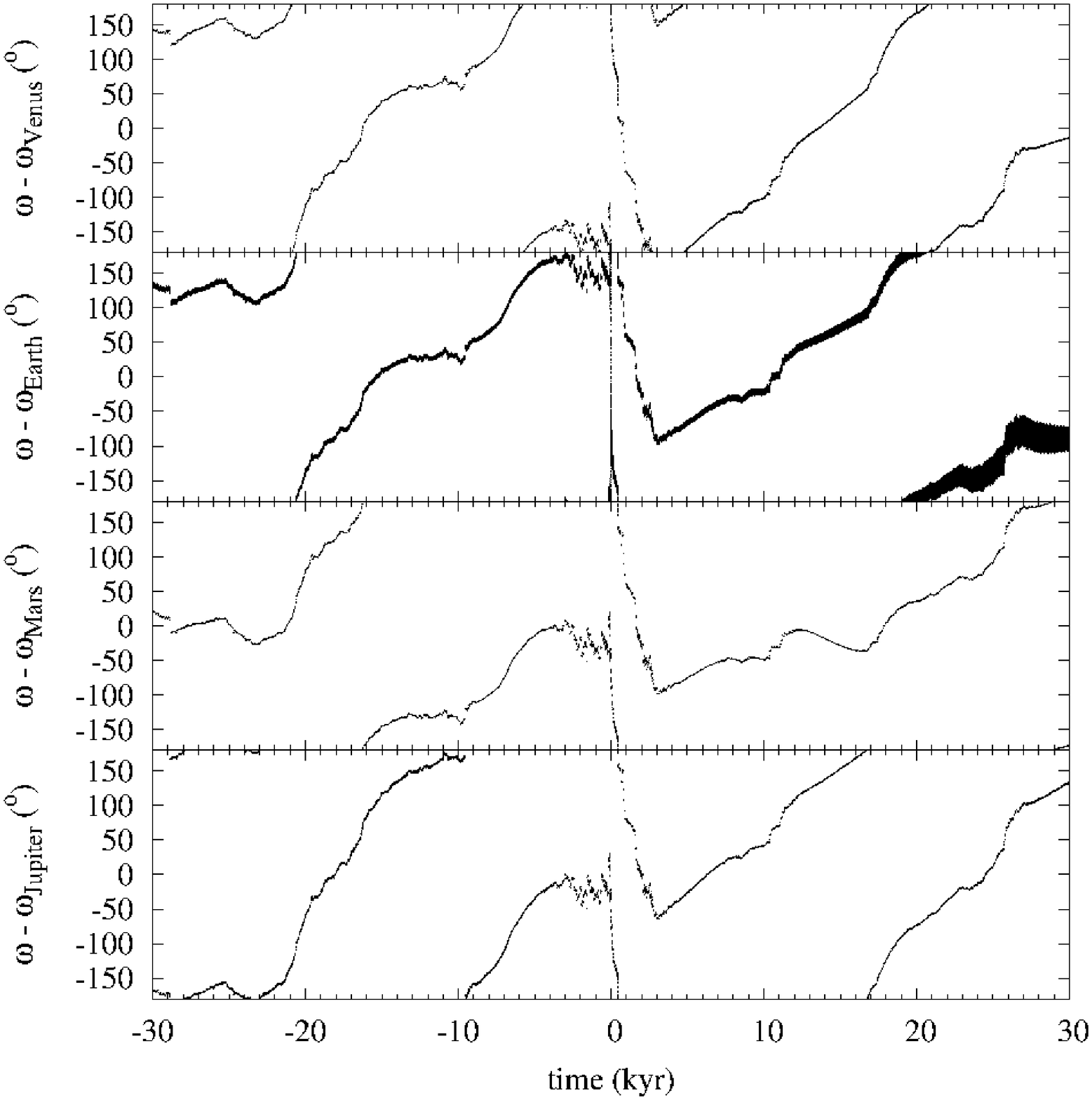}
         \caption{Same as Fig. \ref{fEWfc71T} but for 2014~UR.
                 }
         \label{fEWurT}
      \end{figure}
      }
%
%---------------------------------------------------------------------------------------------------------------------------------
%

      Jupiter is not the single dominant factor regarding the Kozai resonance affecting these objects. The argument of perihelion 
      relative to Jupiter of 2012~FC$_{71}$ was librating around 180\degr\ from 120 to 60 kyr ago and around 90\degr\ for nearly 
      30 kyr soon afterwards, but in general the evolution is very irregular with no persistent periodicities (see Fig. 
      \ref{fEWfc71T}, available electronically only). Less irregular is the equivalent evolution of 2014~EK$_{24}$; in fact, it is 
      the most stable of the four in the $e_{\rm r} \omega_{\rm r}$-portrait. For this object, Jupiter is dominant in the time 
      frame $-11$ to 6 kyr with its relative argument of perihelion librating around 90\degr (see Fig. \ref{fEWek24T}, available 
      electronically only); during the same time interval the argument of perihelion relative to Venus librates around $-90\degr$ 
      and the one relative to Earth flips from $-90\degr$ to 90\degr. During libration episodes, we observe intersecting closed 
      cycles in the $e_{\rm r} \omega_{\rm r}$-plane that resemble ripples in a pond. Mars only seems to affect 2014~EK$_{24}$, a 
      shifting oscillation from $-10$ to $-5$ kyr in Fig. \ref{fEWek24T}. The remarkable periodic behaviour often observed, but 
      without libration, may be understood as resulting from being very close to the separatrix in the phase space of the mean 
      motion resonance. For 2012~FC$_{71}$ there is no actual stable libration of the relative argument of perihelion with respect 
      to any planet in Fig. \ref{fEWfc71T} during the Kozai episode.

      In sharp contrast, the evolution displayed by 2014~QD$_{364}$ in the $e_{\rm r} \omega_{\rm r}$-plane (see Fig. 
      \ref{fEWqd364T}, available electronically only) is very irregular, although multiple but short libration episodes can be 
      identified. The evolution of 2014~UR is cleaner (see Fig. \ref{fEWurT}, available electronically only), but the relative 
      arguments of perihelion hardly circulate. In any case, the superposition of secular resonances makes them more unstable. 
      Short episodes of simultaneous libration break the Kozai state in the case of these objects. The average relative 
      eccentricity with respect to the Earth is also higher for them. The average orbital inclination of 2014~QD$_{364}$ is the 
      lowest, which further explains the observed differences with respect to 2012~FC$_{71}$.

      It is well known that chaos arises where resonances overlap. Overlapping secular resonances have been identified as the 
      source of long-term chaos in the orbits of the terrestrial planets (Lecar et al. 2001; Morbidelli 2002; Lithwick \& Wu 
      2014). The topic of overlapping secular resonances and its effects on the dynamics of asteroids was first studied by Michel 
      (1997) in the particular case of objects moving in Venus horseshoe orbits. He concluded that overlapping of secular 
      resonances is possible, complicating the dynamics of horseshoe orbits significantly. Furthermore, this author found that the 
      presence of overlapping secular resonances provides a transport mechanism to induce high inclinations, switching orbits 
      between different regimes of the Kozai resonance. On a more practical and pragmatic note, overlapping secular resonances
      have a significant impact on active debris removal and the design of disposal strategies for deactivated artificial 
      satellites (Rosengren et al. 2015).

      In our case, Figs. \ref{fEWfc71M}--\ref{fEWurT} show that it is certain that the overlapping of multiple secular resonances 
      ---with Venus, Earth, Mars, and Jupiter--- plays an important role, perhaps dominant, for these four objects and, in 
      general, for those moving within the Arjuna orbital domain (see de la Fuente Marcos \& de la Fuente Marcos 2013). Such
      interplay affects the long-term stability of this type of orbits.

      Ito \& Tanikawa (1999) argued that the terrestrial planets share the effect of the secular perturbation from Jupiter. In 
      particular, the Earth shares the increase in eccentricity of Venus on the long term; the two planets exchange angular 
      momentum (Ito \& Tanikawa 2002). These authors stated that the terrestrial planets maintain their stability by sharing and 
      weakening the secular perturbation from Jupiter (Jupiter's perihelion has a periodicity of nearly 300 kyr); i.e. the 
      Earth--Venus pair distributes the effects of this perturbation. Tanikawa \& Ito (2007) further extended this analysis 
      concluding that, regarding the secular perturbation from Jupiter, the inner planets are a planetary group or collection of 
      loosely connected mutually dynamically dependent planets. This planetary grouping has direct implications on the case 
      studied here; if Jupiter is removed from the calculations, the overlapping secular resonances disappear. Removing the 
      Earth--Moon system or Venus also breaks (but only slightly) the status quo. Only the removal of Mars has virtually 
      negligible effects on the dynamics of the objects discussed in this work. Within this context, objects part of the Arjuna 
      orbital domain are indeed a very singular NEO group. Under certain circumstances, 2012~FC$_{71}$-like orbits, the secular 
      perturbations of Jupiter become negligible (see Fig. \ref{fEWfc71T}, bottom panel) and relatively long-term stability could 
      be achieved.

      Namouni (1999) opened a window to the study of the Kozai realm within the framework of the restricted elliptic three-body 
      problem. Unfortunately, the orbital architecture of the terrestrial zone is significantly more complicated with strong
      secular coupling between planets inducing overlapping of multiple secular resonances. The ultimate source of the 
      perturbation is Jupiter, but that perturbation is shared and weakened by the terrestrial planets.        

   \section{Comparative dynamics and discussion}
      First we will focus on 2012~FC$_{71}$ and 2014~EK$_{24}$, the most stable minor bodies in the set of four discussed here.
      Asteroids 2012~FC$_{71}$ and 2014~EK$_{24}$ represent two incarnations of the same paradigm, passing orbits with small 
      Jacobi constants as described by Namouni (1999). Asteroid 2012~FC$_{71}$ is an Aten and occupies an orbit interior (for the 
      most part) to that of the Earth; in contrast, 2014~EK$_{24}$ is an Apollo that follows an orbit exterior (for the most part) 
      to that of the Earth, yet both currently exhibit the same secular resonance with libration of $\omega$ around 0\degr. The 
      orbit of 2012~FC$_{71}$ is far more stable than that of 2014~EK$_{24}$, however. Asteroid 2012~FC$_{71}$ stays in the Kozai 
      resonance for at least 330 kyr, but 2014~EK$_{24}$ only remains there for 10 to 30 kyr, suggesting that Apollo passing 
      companions are far less stable than Aten ones. Our calculations show that the stable island occupied by Aten asteroid 
      2012~FC$_{71}$ is less perturbed and its volume in orbital parameter space is much larger than the one hosting Apollo 
      asteroid 2014~EK$_{24}$. The near mean motion resonances and secular resonances discussed above are clearly stronger for the 
      stable Apollo island. The key to understanding this is in the behaviour of the argument of perihelion relative to Mars: in 
      Fig. \ref{fEWfc71T} it circulates rapidly, but in Fig. \ref{fEWek24T} an almost proper libration around 90\degr\ for nearly 
      5 kyr is observed. The effect of the secular perturbations linked to near resonances significantly reduces the stability of 
      the Apollo island with respect to that of the Aten island.  

      Unlike  horseshoe librators, NEOs trapped in the Kozai resonance have a very slow orbital evolution (see Figs. 
      \ref{controlfc71}, \ref{controlfc712}, and \ref{controlek24}) and can remain relatively unperturbed for hundreds of 
      thousands of years (Michel \& Thomas 1996; Gronchi \& Milani 1999) as they never get excessively close to the Earth--Moon 
      system (see panel A in Figs. \ref{controlfc71}, \ref{controlfc712}, and \ref{controlek24}), well beyond the Hill radius of 
      our planet (0.0098 AU) during the span of the passing co-orbital phase. They do not librate in $\lambda_{\rm r}$; in other 
      words, they are not classical co-orbitals like Trojans, quasi-satellites, or horseshoe librators. However, their paths 
      resemble those of the classical co-orbitals (see Figs. 1 and 5 in de la Fuente Marcos \& de la Fuente Marcos 2013). For 
      these objects, close encounters below the Hill radius are observed prior to their insertion in the Kozai resonance and 
      immediately before their eviction from it. Critical close encounters for 2012~FC$_{71}$ always happen at the descending 
      node, but for 2014~EK$_{24}$ they take place at both nodes (see panel G in Figs. \ref{controlfc71}, \ref{controlfc712}, and 
      \ref{controlek24}). In most of the control orbits of 2014~EK$_{24}$, the object was a horseshoe librator to the Earth before 
      becoming a passing co-orbital; in some cases it returns to that state right after leaving the Kozai resonance. Although not 
      shown, during the passing co-orbital episodes $\dot{\Omega}_{\rm r} < 0$ in agreement with predictions made in Namouni 
      (1999). 

      Asteroids 2014 QD$_{364}$ and 2014 UR are clearly more unstable than the previous two (see Appendix A of the online 
      materials) and this is not the result of having more uncertain orbits. The current orbital solution of 2014 UR is as robust 
      as that of 2014 EK$_{24}$, but the object is far less stable. Both 2014 QD$_{364}$ and 2014 UR are likely smaller than the 
      other two, they could be fragments of fragments. The current catastrophic disruption rate of asteroids in the main belt has 
      been recently studied by Denneau et al. (2015). These authors have found that the rate is significantly higher than 
      previously thought with rotational disruptions being the dominant source of most fragments. Asteroidal decay could be 
      induced by collisional processes (see e.g. Dorschner 1974; Ryan 2000), but could also be the combined result of thermal 
      fatigue (see e.g. \v{C}apek \& Vokrouhlick\'y 2010) and rotational (see e.g. Walsh et al. 2008) or tidal stresses (see e.g. 
      T\'oth et al. 2011). These last three processes can easily produce secondary fragments. Planetary ejecta from Mars, the 
      Earth--Moon system, and Venus are also possible sources (Warren 1994; Gladman et al. 1995; Bottke et al. 1996; Gladman 1996, 
      1997). Asteroid 2014~EK$_{24}$ is a fast rotator (see above) and the largest of the group. From a theoretical standpoint, it 
      may be producing additional fragments via rotational stress.  These putative fragments will follow similar orbits, at least 
      initially. Asteroid 2013 RZ$_{53}$ (see Table \ref{members}) follows a very similar orbit and it has a size in the range 
      1--4 m, typical of a fragment of a fragment.

   \section{Artificial interlopers}
      The four objects studied here are part of a subclass of a larger group of dynamically cold, resonant asteroids known as the 
      Arjunas (see de la Fuente Marcos \& de la Fuente Marcos 2013, 2015). Nearly every time a new object moving in a very 
      Earth-like orbit is discovered, it is speculated that it may be a relic of human space exploration which has returned to the
      neighbourhood of the Earth--Moon system. Items of space debris are routinely found when looking for asteroids. Spectral and 
      photometric observations, and orbital analysis concluded that J002E3, initially considered a new minor body, was an Apollo 
      rocket body (Jorgensen et al. 2003). The Rosetta spacecraft was also mistaken as an asteroid and given the designation 
      2007~VN$_{84}$ (Kowalski et al. 2007).\footnote{\url{http://www.minorplanetcenter.net/iau/mpec/K07/K07V70.html}} Even the
      European Space Agency telescope Gaia was erroneously given a minor body designation, 2015~HP$_{116}$ (Anderson et al. 
      2015).\footnote{\url{http://www.minorplanetcenter.net/iau/mpec/K15/K15HC3.html}}\footnote{\url{http://www.minorplanetcenter.net/iau/mpec/K15/K15HC5.html}}

      Therefore, within the Arjuna orbital domain sit artificial interlopers, objects that left low-Earth parking orbits and 
      headed out to interplanetary space. A recent example is 2014 XX$_{39}$, which has been identified as hardware from the 
      Hayabusa 2 mission, specifically the Hayabusa 2 spacecraft. Hayabusa 2 is an asteroid sample return mission aimed at the 
      C-type asteroid 162173 (1999~JU$_{3}$) (Tsuda et al. 2013; Giancotti et al. 2014). This recent misidentification and the 
      ones pointed out above highlight the fact that the positional information available on distant artificial objects is 
      incomplete. Table \ref{XX39} (available electronically only) shows the heliocentric Keplerian orbital elements of 
      2014 XX$_{39}$ before it was identified as the Hayabusa 2 spacecraft. Its orbit is quite uncertain because it was observed 
      just six times after its discovery on 2014 December 6. Figure \ref{controlxx39} (available electronically only) shows the 
      evolution of this object as well as that of the nominal orbit of the actual Hayabusa 2 spacecraft (data from the JPL 
      HORIZONS system). Hayabusa 2 was launched on December 3, therefore the orbital evolution prior to that date is only 
      displayed here as a mere dynamical curiosity. The aim of our analysis is to compare the orbital evolution of a confirmed 
      human-made piece of hardware to that of natural objects like 2012~FC$_{71}$, 2014~EK$_{24}$, 2014~QD$_{364}$, or 2014~UR. 
      For both orbits displayed in Fig. \ref{controlxx39} the overall evolution is different from that in Figs. 
      \ref{controlfc71}--\ref{controlur}; the object tends to drift away from Earth's co-orbital region rather quickly. The launch 
      of a spacecraft has to be very fine tuned so its subsequent dynamical evolution resembles that of a natural object like the 
      Arjunas. This is not surprising taking into account that the phase space in the neighbourhood of our planet is threaded by a 
      dense stochastic web of instabilities.  
%
%------------------------------------------------------------------------------------------------------------------------- TABLE I
%--------------------------------------------------------------------------------------------- Orbital elements asteroid 2014 XX39
%
     \onltab{
     \begin{table}
      \fontsize{8}{11pt}\selectfont
      \tabcolsep 0.15truecm
      \caption{\label{XX39}Heliocentric Keplerian orbital elements of 2014 XX$_{39}$. 
              }
      \begin{tabular}{ccc}
       \hline\hline
        Parameter                                         &   &   Value             \\
       \hline
        Semi-major axis, $a$ (AU)                          & = &   1.0023$\pm$0.0004 \\
        Eccentricity, $e$                                 & = &   0.088$\pm$0.003   \\
        Inclination, $i$ (\degr)                          & = &   6.9$\pm$0.2       \\
        Longitude of the ascending node, $\Omega$ (\degr) & = & 250.65$\pm$0.12     \\
        Argument of perihelion, $\omega$ (\degr)          & = &  95.8$\pm$0.3       \\
        Mean anomaly, $M$ (\degr)                         & = &  80.0$\pm$0.2       \\
        Perihelion, $q$ (AU)                              & = &   0.914$\pm$0.003   \\
        Aphelion, $Q$ (AU)                                & = &   1.0905$\pm$0.0005 \\
        Absolute magnitude, $H$ (mag)                     & = &  26.6               \\
       \hline
      \end{tabular}
      \tablefoot{Values include the 1$\sigma$ uncertainty. The orbit is computed at Epoch JD 2457000.5 that corresponds to 
                 0:00 UT on 2014 December 9 (J2000.0 ecliptic and equinox). Source: JPL Small-Body Database.
                }
     \end{table}
     }
%
%---------------------------------------------------------------------------------------------------------------------------------
%
%
%---------------------------------------------------------------------------------------------------------------------------------
%
      \onlfig{
      \begin{figure*}
        \centering
         \includegraphics[width=\linewidth]{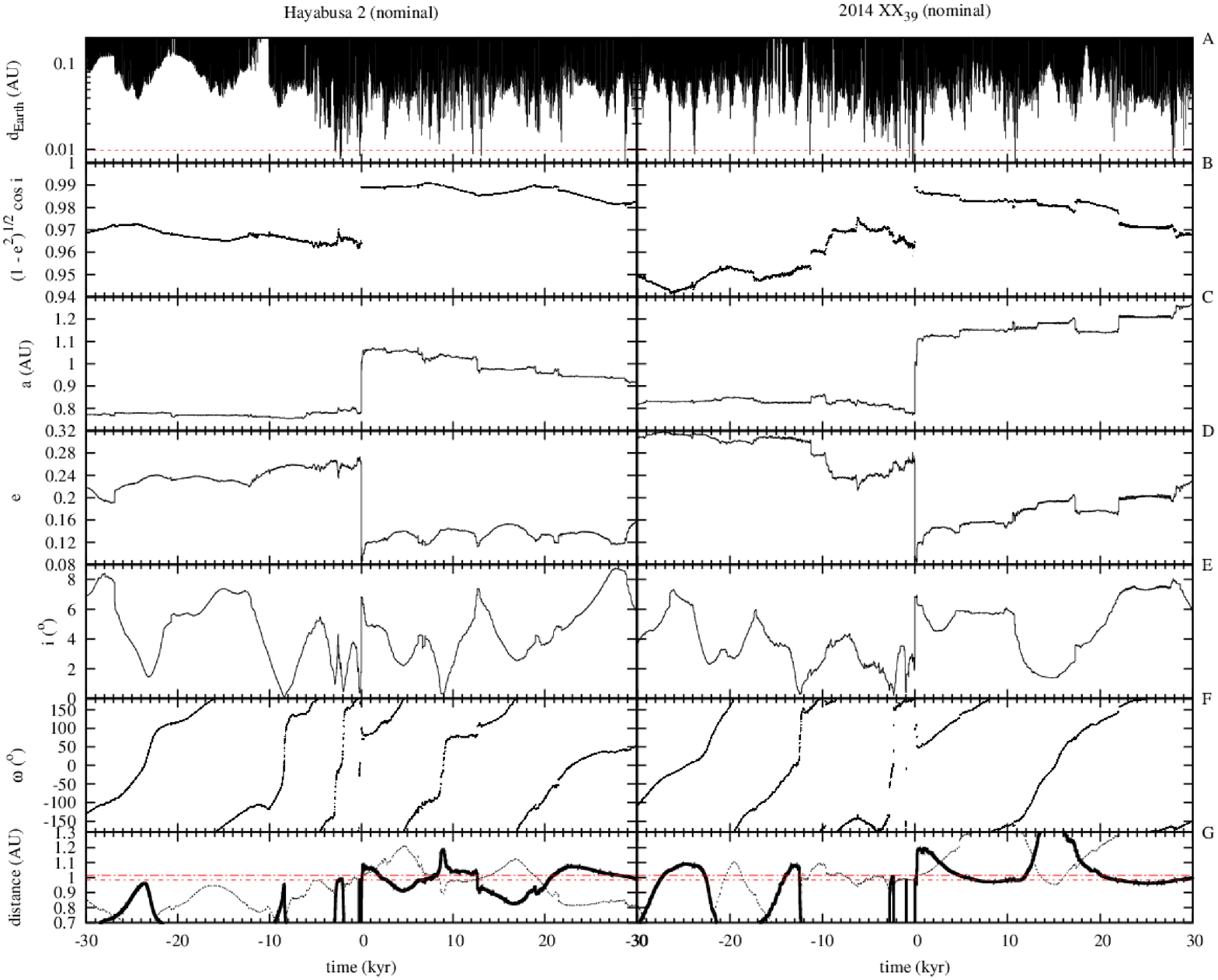}
         \caption{Same as Fig. \ref{controlfc71} but for 2014~XX$_{39}$ and Hayabusa 2 (see  text for details).
                 }
         \label{controlxx39}
      \end{figure*}
      }
%
%---------------------------------------------------------------------------------------------------------------------------------
%
 
      In the case of known objects moving in Arjuna-type orbits, we can easily conclude that they are unlikely to be human-made. 
      The first unambiguously documented objects sent out of the atmosphere were launched in the second half of 1957. Objects in 
      Table \ref{members} were all discovered near their most recent perigee; their previous perigee appear listed in the table 
      under column $L$. Only four objects had a previous perigee after the second half of 1957: 2006 JY$_{26}$, 2006 RH$_{120}$, 
      2012 FC$_{71}$, and 2012 LA$_{11}$. With the exception of 2006 RH$_{120}$, none of their perigees can be matched to any known
      rocket launch. Asteroid 2006 RH$_{120}$ reached perigee on 1979 May 15, the same day the Russian satellite Kosmos 1098 was 
      launched. However, this was just a coincidence because 2006 RH$_{120}$ is a confirmed natural object (Kwiatkowski et al. 
      2009; Granvik et al. 2012). None of the objects in Table \ref{members} are expected to be human-made.
%
%----------------------------------------------------------------------------------------------------------------------- TABLE III
%
       \begin{table*}
        \centering
        \fontsize{8}{11pt}\selectfont
        \tabcolsep 0.20truecm
        \caption{\label{members}Orbital properties of objects currently moving in Arjuna-type orbits.  
                }
        \begin{tabular}{ccccccccccc}
         \hline\hline
           Object          &    $a$   &   $e$    &   $i$   &  $\Omega$ &  $\omega$ &  MOID    & Class  &   $S$   &     $L$     & $H$   \\
                           &   (AU)   &          & (\degr) &   (\degr) &   (\degr) &  (AU)    &        &   (yr)  &             & (mag) \\
         \hline
           2003 YN$_{107}$ & 0.988674 & 0.013957 & 4.32109 & 264.42634 &  87.58224 & 0.004774 & Aten   &  56.655 & 1938-Jul-10 & 26.50 \\
           2006 JY$_{26}$  & 1.010089 & 0.083069 & 1.43923 &  43.47088 & 273.60219 & 0.000395 & Apollo &  68.841 & 1971-Dec-09 & 28.40 \\
           2006 RH$_{120}$ & 0.998625 & 0.019833 & 1.52613 & 290.52215 & 177.87923 & 0.000679 & Aten   & 402.527 & 1979-May-15 & 29.50 \\
           2008 KT         & 1.010871 & 0.084831 & 1.98424 & 240.62771 & 102.09585 & 0.000462 & Apollo &  63.816 & 1952-Nov-29 & 28.20 \\
           2008 UC$_{202}$ & 1.010329 & 0.068598 & 7.45257 &  37.35244 &  91.72231 & 0.001714 & Apollo &  67.213 & 1946-Apr-29 & 28.30 \\
           2009 BD         & 1.008614 & 0.040818 & 0.38516 &  58.48799 & 110.50392 & 0.003565 & Apollo &  80.872 & 1955-Jul-16 & 28.10 \\
           2009 SH$_{2}$   & 0.991559 & 0.094222 & 6.81115 &   6.69814 & 101.64342 & 0.000762 & Aten   &  75.673 & 1957-Mar-27 & 24.90 \\
           2010 HW$_{20}$  & 1.010993 & 0.050062 & 8.18472 &  39.23393 &  60.25567 & 0.007961 & Apollo &  63.099 & 1952-Apr-23 & 26.10 \\
           2012 FC$_{71}$  & 0.988482 & 0.088006 & 4.94336 &  38.18431 & 348.04099 & 0.057787 & Aten   &  55.718 & 1959-Jun-01 & 25.20 \\
           2012 LA$_{11}$  & 0.988483 & 0.096639 & 5.11877 & 260.69449 & 241.18460 & 0.009692 & Aten   &  55.725 & 1980-Jun-09 & 26.10 \\
           2013 BS$_{45}$  & 0.993678 & 0.083875 & 0.77337 &  83.55082 & 149.70624 & 0.011479 & Aten   & 100.236 & 1934-Aug-16 & 25.90 \\
           2013 RZ$_{53}$  & 1.013056 & 0.031689 & 2.09522 & 345.60077 &  68.42060 & 0.005644 & Apollo &  53.047 & 1954-Mar-13 & 31.10 \\
           2014 EK$_{24}$  & 1.004322 & 0.072327 & 4.72207 & 341.91526 &  62.44932 & 0.034374 & Apollo & 165.836 & 1933-Feb-22 & 23.20 \\
           2014 QD$_{364}$ & 0.989075 & 0.041224 & 3.97051 & 158.24462 &  28.35946 & 0.014553 & Aten   &  58.712 & 1955-Aug-30 & 27.20 \\
           2014 UR         & 0.999052 & 0.013243 & 8.22675 &  25.33163 & 247.50567 & 0.008054 & Aten   & 542.993 & 1940-Oct-20 & 26.60 \\
         \hline
        \end{tabular}
        \tablefoot{The orbital elements have been computed at Epoch 2457000.5 (2014 December 9) with the exception of 
                   2006~RH$_{120}$ (2454115.5) and 2009~BD (2455200.5). $S$ is the synodic period. $L$ is the date of the closest 
                   approach prior to discovery. Source: JPL Small-Body Database. Data as of 2015 May 13.
                  } 
       \end{table*} 
%
%---------------------------------------------------------------------------------------------------------------------------------
%

   \section{Conclusions}
      Our calculations confirm that, in accordance with predictions made by Namouni (1999), minor bodies moving in passing orbits 
      with a small Jacobi constant relative to a host planet, or co-orbital passing orbits, do exist. Asteroids 2012~FC$_{71}$, 
      2014~EK$_{24}$, 2014~QD$_{364}$, and 2014~UR exhibit an orbital evolution unlike any other known NEO; however, they are all 
      part of a little studied group of dynamically cold, resonant asteroids known as the Arjunas (see de la Fuente Marcos \& de 
      la Fuente Marcos 2013, 2015) that occupy the most dynamically stable region of the NEO orbital parameter space. This region 
      is affected by multiple mean motion, near mean motion, and secular resonances. The superposition of this relatively large 
      set of resonances contributes to both stabilising and destabilising the trajectories of the bodies moving within this 
      peculiar orbital parameter space. The four objects studied in this work are submitted to Kozai resonances. 

      They are small objects that can only be observed when they arrive close to the Earth, with a frequency of many decades or 
      even centuries, remaining within 0.5 AU from the Earth for about 10--20 yr with typical close approaches at minimum 
      distances of 0.05 AU. This implies that the vast majority of them may be well beyond reach of current NEO surveys (see the 
      analysis in de la Fuente Marcos \& de la Fuente Marcos 2015); however, they could be numerous because they are dynamically 
      stable (in particular the Atens with orbits similar to that of 2012~FC$_{71}$), significantly more stable than typical NEOs 
      or even Earth quasi-satellites like 2013~LX$_{28}$ (Connors 2014). Most known Earth quasi-satellites also exhibit Kozai-like 
      dynamics (de la Fuente Marcos \& de la Fuente Marcos 2014). Asteroid 2014~EK$_{24}$ represents a rare instance of an object 
      that is in very near mean motion resonance with multiple planets (Venus, Earth, Mars, and Jupiter). The chance discovery of 
      three objects with similar dynamical properties within a few months during 2014 strongly suggests that the size of this 
      population may not be as negligible as considered in typical studies (see e.g. Rabinowitz et al. 1993; Bottke et al. 2002; 
      Brasser \& Wiegert 2008).

      As members of the unofficial Arjuna class, these minor bodies moving in orbits with low-eccentricity, low-inclination, and 
      Earth-like period eventually suffer a close encounter with the Earth--Moon system that changes their dynamical status, 
      turning them into traditional co-orbitals of the horseshoe, Trojan, or quasi-satellite type, or ---more often--- ejecting 
      them outside Earth's co-orbital region. Objects of this dynamical class may experience repeated co-orbital and Kozai 
      episodes, and transitions are driven by the encounters (de la Fuente Marcos \& de la Fuente Marcos 2013). The most usual 
      (and persistent) state is symmetric horseshoe, but quasi-satellite and Trojan episodes have also been observed during the 
      simulations. Prior to the Kozai episode and after it, these minor bodies often follow horseshoe trajectories. When trapped 
      in the Kozai resonant state, these objects exhibit libration of their arguments of perihelion around 0\degr\ or 180\degr, 
      characteristic of passing orbits with small Jacobi constants. 

      Although NEOs in this group can remain orbitally stable for many thousands of years, their secular dynamics are quite 
      complex and cannot be properly described within the framework of the three-body problem alone. The orbital evolution of 
      these objects, and the Arjuna asteroids in general, is organised to a large extent by secular chaos. From an observational 
      point of view, minor bodies in this group are amongst the most dramatically restricted NEOs regarding favourable visibility 
      windows because these are separated in time by many decades or even several centuries. Failure to secure more observations 
      of 2012~FC$_{71}$ or 2014~QD$_{364}$ before they leave Earth's neighbourhood in the near future can definitely place them in 
      the category of ``lost asteroids'' because their next visibility window comes half a century from now.

   \begin{acknowledgements}
      The authors thank the referee for his/her constructive report and very helpful suggestions regarding the presentation of 
      this paper. The authors would like to thank S. J. Aarseth for providing the code used in this research, and O. Vaduvescu and 
      R. L. Cornea for sharing their results on 2014~EK$_{24}$ prior to publication. This work was partially supported by the 
      Spanish `Comunidad de Madrid' under grant CAM S2009/ESP-1496. Some of the calculations presented here were completed on the 
      `Servidor Central de C\'alculo' of the Universidad Complutense de Madrid. In preparation of this paper, we made use of the 
      NASA Astrophysics Data System, the ASTRO-PH e-print server, the MPC data server, and the NEODyS information service.      
   \end{acknowledgements}
 
   \bibliographystyle{aa}

  \Online

  \begin{appendix}
     \section{Average short-term evolution of 2012~FC$_{71}$, 2014~EK$_{24}$, 2014~QD$_{364}$, and 2014~UR}
\begin{figure}
        \centering
         \includegraphics[width=\linewidth]{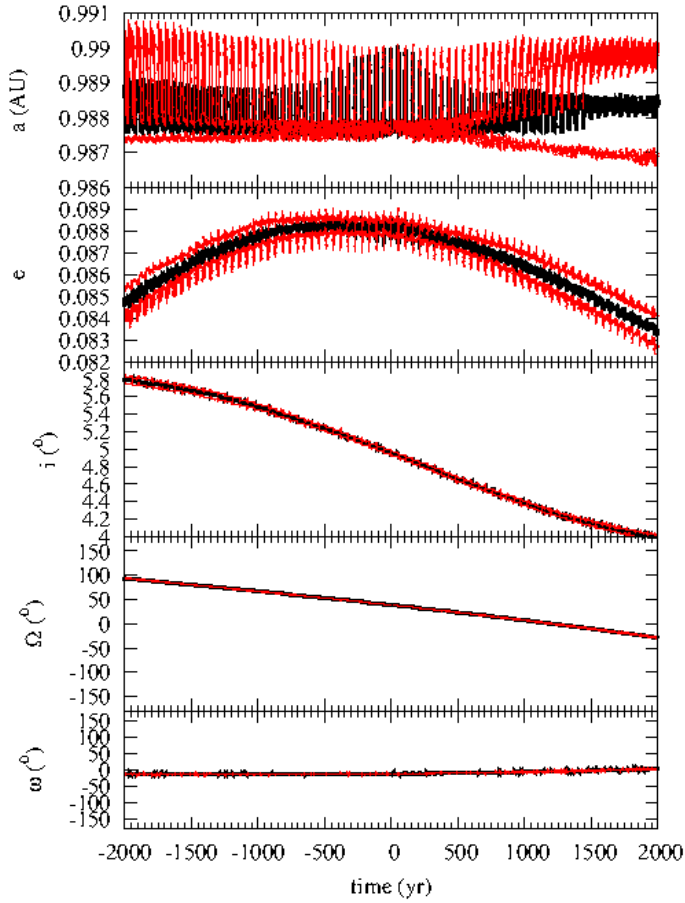}
         \caption{Time evolution of the orbital elements $a$, $e$, $i$, $\Omega$, and $\omega$ of 2012~FC$_{71}$. The thick curve 
                  shows the average evolution of 100 control orbits, the thin curves show the ranges in the values of the 
                  parameters at the given time.
                 }
         \label{STEfc71}
      \end{figure}

\begin{figure}
        \centering
         \includegraphics[width=\linewidth]{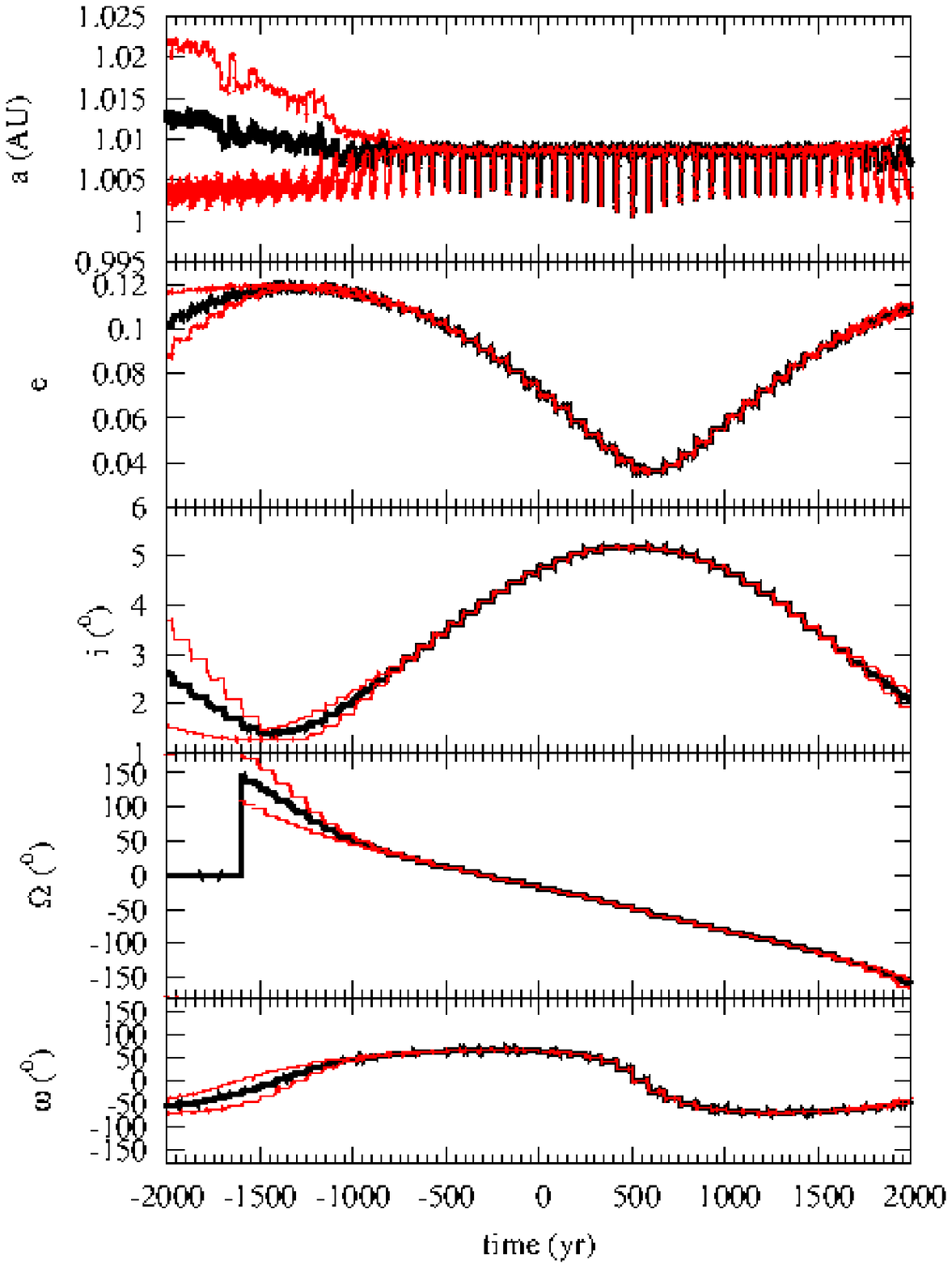}
         \caption{Same as Fig. \ref{STEfc71} but for 2014~EK$_{24}$.
                 }
         \label{STEek24}
      \end{figure}

\begin{figure}
        \centering
         \includegraphics[width=\linewidth]{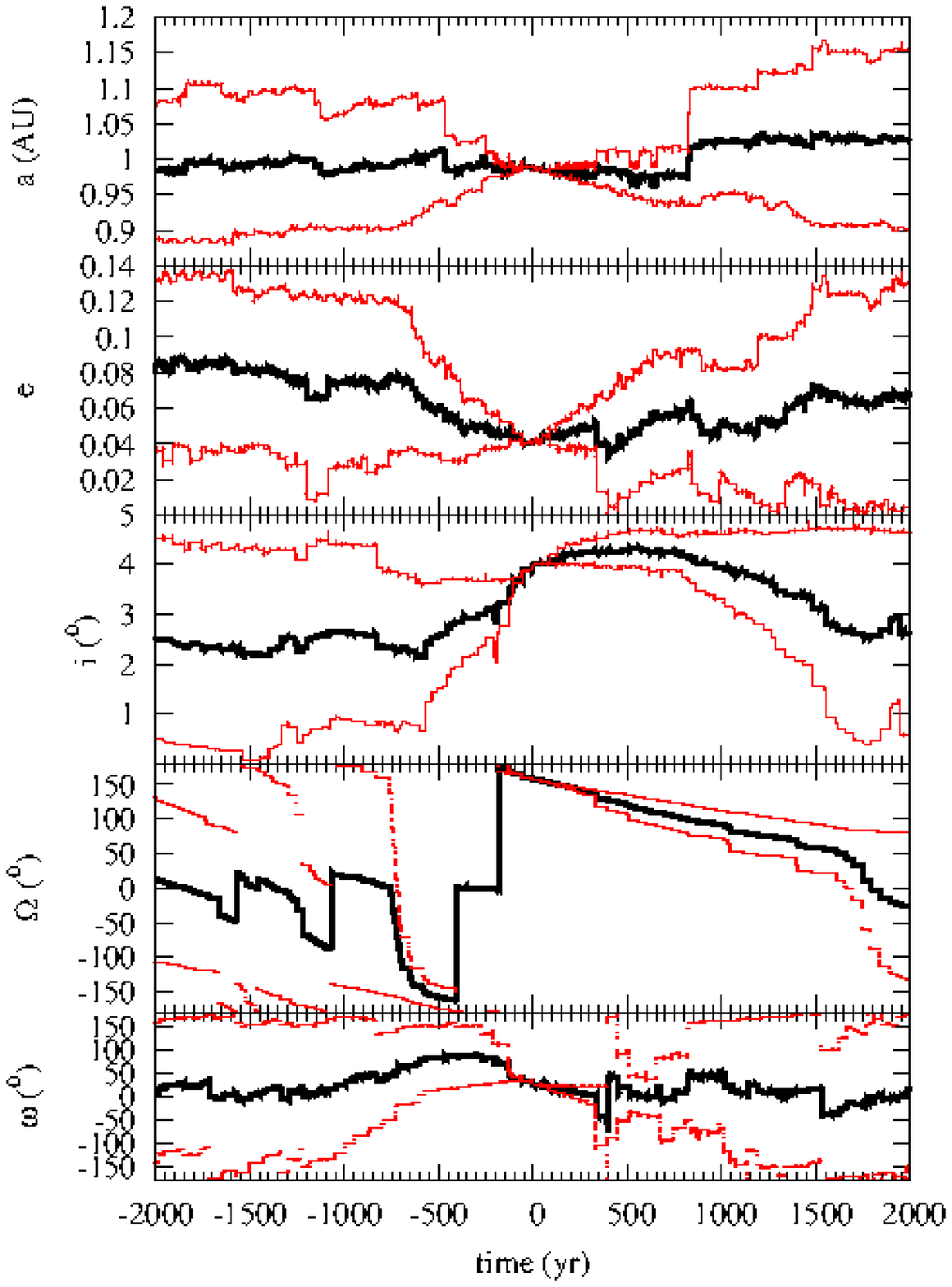}
         \caption{Same as Fig. \ref{STEfc71} but for 2014~QD$_{364}$.
                 }
         \label{STEqd364}
      \end{figure}

\begin{figure}
        \centering
         \includegraphics[width=\linewidth]{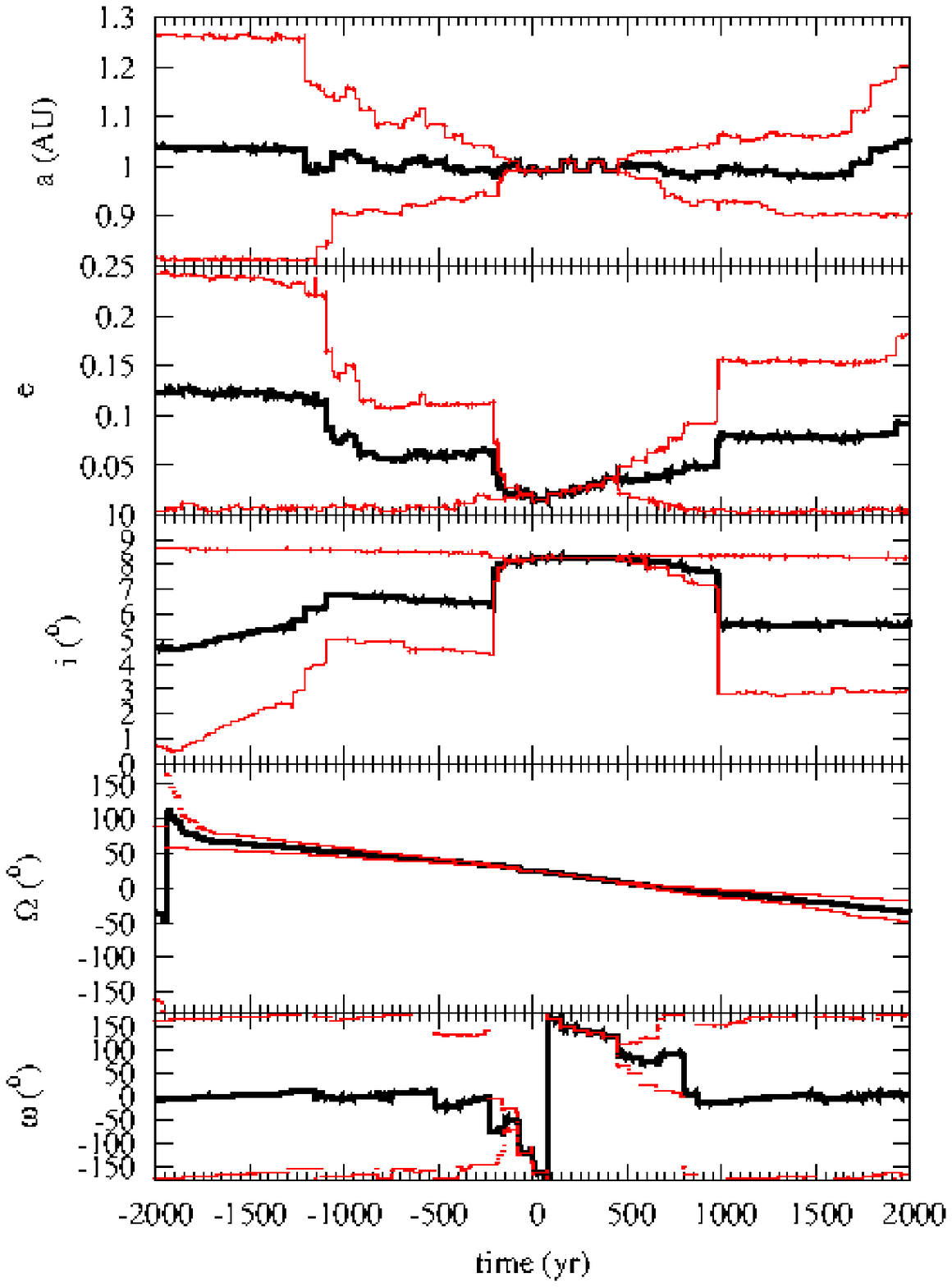}
         \caption{Same as Fig. \ref{STEfc71} but for 2014~UR.
                 }
         \label{STEur}
      \end{figure}

        Figures \ref{STEfc71} to \ref{STEur} show the short-term evolution of the orbital elements $a$, $e$, $i$, $\Omega$, and 
        $\omega$ of the objects studied here. The thick black curves show the average results of the evolution of 100 control
        orbits computed as described in Section 2. The thin red curves show the ranges (minimum and maximum) in the values of the 
        parameters at a given time. Asteroid 2012~FC$_{71}$ is the most dynamically stable of the four objects, followed by 
        2014~EK$_{24}$; 2014~UR is quite unstable and the least stable is 2014~QD$_{364}$. Asteroid 2014~QD$_{364}$ has an 
        $e$-folding time, or characteristic timescale on which two arbitrarily close orbits diverge exponentially, of a few dozen 
        years. In sharp contrast, the $e$-folding time of 2014~EK$_{24}$ is several hundred years. Surprisingly, the $e$-folding 
        time of 2014~EK$_{24}$ is somewhat longer than that of 2012~FC$_{71}$, even if its orbital evolution is significantly less 
        stable.
%
%---------------------------------------------------------------------------------------------------------------------------------
%
%
%---------------------------------------------------------------------------------------------------------------------------------
%
%
%---------------------------------------------------------------------------------------------------------------------------------
%
%
%---------------------------------------------------------------------------------------------------------------------------------
%
%
%---------------------------------------------------------------------------------------------------------------------------------
%
%
%---------------------------------------------------------------------------------------------------------------------------------
%
%
%---------------------------------------------------------------------------------------------------------------------------------
%
%
%---------------------------------------------------------------------------------------------------------------------------------
%
  \end{appendix}

\end{document}